\documentclass[
 aip,
 amsmath,amssymb,
 cha,
 reprint,%
]{revtex4-1}
\usepackage[caption=false]{subfig}
\usepackage{amssymb}
\usepackage{bm}
\usepackage{amsmath}
\usepackage{xcolor}
\usepackage{mathtools}
\usepackage[normalem]{ulem}
\usepackage{mathrsfs}
\newcommand{\ie}{\textit{i.e.}}
\newcommand{\D}{{\rm d}}
\newcommand{\img}[0]{\textrm{i}}
\newcommand{\etal}[0]{\textit{et al.}}
\usepackage{mathptmx}
\usepackage{etoolbox}

\makeatletter
\def\@email#1#2{%
 \endgroup
 \patchcmd{\titleblock@produce}
  {\frontmatter@RRAPformat}
  {\frontmatter@RRAPformat{\produce@RRAP{*#1\href{mailto:#2}{#2}}}\frontmatter@RRAPformat}
  {}{}
}%
\makeatother
\begin{document}

\preprint{AIP/123-QED}

\title[Synchronization transitions in ensembles of active particles with distance-dependent higher-order interactions]{Synchronization transitions in ensembles of active particles with distance-dependent higher-order interactions}
\author{Dhrubajyoti Biswas}
\affiliation{Indian Institute of Technology Kharagpur, West Medinipur, West Bengal, 721302, India}
\email{dhrubajyoti@phy.iitkgp.ac.in, dhrubajyoti98@gmail.com; Corresponding Author.}

\author{P.S. Burada}
\affiliation{Indian Institute of Technology Kharagpur, West Medinipur, West Bengal, 721302, India}
\email{psburada@phy.iitkgp.ac.in}

\author{Pintu Patra}
\affiliation{Indian Institute of Technology Kharagpur, West Medinipur, West Bengal, 721302, India}
\email{pintupatra@phy.iitkgp.ac.in}

\date{\today}

\begin{abstract}
Higher-order interactions are ubiquitous in biological systems and can be used to better describe the ensemble dynamics of self-propelled particles.
To this end, this article investigates a Vicsek-Kuramoto model with distance-dependent higher-order interactions and obtains analytical results, albeit under certain assumptions, that characterize its collective dynamics. Subsequently, these are validated through simulations of the system equations, demonstrating both continuous and explosive transitions between the asynchronous and synchronous collective states. Further numerical results also reveal that short-range interactions and tightly clustered initial states can lead to deviations from the theoretical predictions, whereas the latter can be offset by increasing the common self-propulsion velocity. These results offer insights into a broad class of active matter systems with higher-order interactions, and provide a framework for designing and utilizing them for specific applications.
\end{abstract}

\maketitle


\section{Introduction}

\begin{figure}[h]
    \centering
    \subfloat[\label{single_particle}]{\includegraphics[width=0.28\linewidth]{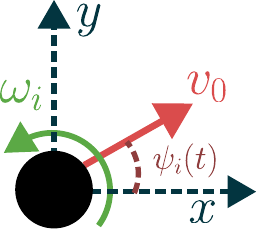}}\hfill
    \subfloat[\label{multiple_particle}]{\includegraphics[width=0.27\linewidth]{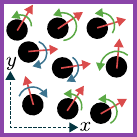}}\hfill
    \subfloat[\label{hoi-network}]{\includegraphics[width=0.27\linewidth]{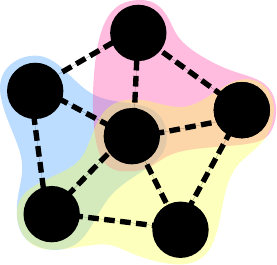}}
    \caption{Schematic representation of a single (in (a)) and an ensemble of chiral active particles (in (b)), respectively, having a common speed $v_0$ and autonomous rotational frequency $\omega_i$ (either clockwise or anticlockwise, indicated by blue and green arrows, respectively). Here, the instantaneous direction of self-propulsion $\psi_i(t)$ is indicated by the red arrow. In (c), a generalized interaction network is demonstrated, where the black lines and coloured regions denote pairwise and higher-order interactions, respectively.}
\end{figure}

The mathematical formalism of dynamical systems, combined with methods from network science, can be used to accurately describe many natural phenomena and to tune artificial systems towards specific applications. With examples ranging from the flocking of birds to the jamming of traffic~\cite{cristiani2021all,PhysRevE.59.4857}, such systems are typically described as a large ensemble of individual agents (or nodes), the dynamics of which are expressed through large sets of coupled difference and differential equations. A key characteristic of such systems is the coupling between internal and spatial degrees of freedom, and emergent collective behavior arises from their intricate interplay. Given their ubiquity, a variety of mathematical models have been proposed to capture different aspects of such physical systems. Among the most influential are the Kuramoto~\cite{RevModPhys.77.137} and Stuart-Landau~\cite{kg1l-8dvc,garcia2012complex} oscillators, which describe coupled oscillators that are typically fixed in space. In contrast, the Vicsek model~\cite{vicsek2012collective,ginelli2016physics} describes the collective motion of active particles whose dynamics are governed by an interplay between self-propulsion, alignment interactions, and fluctuations; see Figs.~\ref{single_particle} and \ref{multiple_particle} for a schematic. More recently, the \textit{swarmalator} model has also emerged as a unified dynamical framework that simultaneously incorporates internal and spatial motion~\cite{o2017oscillators,lizarraga2020synchronization,sar2022dynamics,PhysRevE.105.014211,PhysRevE.98.022203}.

An important feature of a large class of active particles is \textit{chirality}, which characterizes their \textit{intrinsic} tendency to self-rotate~\cite{lowen2016chirality,liebchen2022chiral}, 
in the absence of fluctuations and interparticle coupling. 
It arises naturally in a variety of systems, typically from asymmetries in its physical properties~\cite{patra2022collective,witten2020review}, with examples that range from motile microorganisms~\cite{riedel2005self,lauga2006swimming,erglis2007dynamics,friedrich2007chemotaxis} to chemically active Janus particles~\cite{ao2015diffusion,PhysRevResearch.2.013250} and other synthetically produced active swimmers~\cite{PhysRevLett.110.198302}. Mathematically, chirality is commonly incorporated into the dynamical equations via a nonzero natural frequency in the orientational dynamics, generally sampled from a distribution, yielding heterogeneous populations of clockwise and anticlockwise particles~\cite{wang2024condensation}. The interplay between translation and rotation gives rise to a wealth of dynamical phenomena, including circular~\cite{van2008dynamics} and helical~\cite{4kxb-h6p4} trajectories, vortex formation~\cite{caprini2024self}, clustering~\cite{ma2022dynamical}, and unconventional collective states~\cite{PhysRevLett.119.058002,PhysRevLett.127.238001} that are typically absent in non-chiral systems. However, despite recent progress, understanding the interplay between chirality, non-trivial network structures, and interaction patterns remains a central challenge in the study of active matter particles and their emergent dynamical properties.

While the chirality of individual particles is an important feature that enriches the collective dynamics observed in different models of active matter systems, most existing studies in this domain continue to assume that interactions between the particles are exclusively pairwise. However, recent theoretical studies~\cite{leon2025collectivedynamicshigherordervicsek,marghoti2026impactchanneldynamicshigherorder} have demonstrated that higher-order alignment rules may emerge naturally from microscopic interaction mechanisms, in addition to data-driven techniques~\cite{karmelic2022emergent,niedostatek2025mining,lotito2024exact,PhysRevX.14.011050,cencetti2021temporal,chelaru2021high} that highlight the importance of such interactions. These interactions enable connections among three or more nodes, whose properties cannot be adequately captured by any combination and/or superposition of pairwise interactions. These are, therefore, fundamental units of connectivity in realistic network structures, a simple representative schematic of which is shown in Fig.~\ref{hoi-network}. 
The effect of these higher-order structures have been investigated in a variety of systems, including but not limited to neuronal populations~\cite{parastesh2022synchronization,huang2025effects,farrera2025explosive,biswas2025effect}, biochemical reaction networks~\cite{silva2024graph,restrepo2026higher}, 
ecological communities~\cite{bairey2016high,grilli2017higher,mayfield2017higher,gibbs2022coexistence,gallardo2024higher}, and social dynamics~\cite{alvarez2021evolutionary,xu2023linear,battiston2025higher,yu2026coevolutionary}. Despite these and other such advances (see Majhi~\etal~\cite{majhi2022dynamics} and Battiston~\etal~\cite{battiston2026collective} for recent reviews), higher-order interactions in active matter dynamics remain relatively unexplored due to their inherent complexity, motivating the investigation of collective dynamics in models that extend beyond the conventional pairwise framework.

In this context, the present work explores an analytically tractable Vicsek-Kuramoto model of active particles that accounts for self-propulsion and chirality, as well as distance-dependent pairwise and higher-order interactions. It provides detailed insights, obtained through analytical techniques under suitable approximations, supported by extensive numerical simulations of the system equations. The interplay between higher-order interactions, self-propulsion, and chirality is investigated, with the latter being incorporated via heterogeneous natural frequencies sampled from a Cauchy distribution. In particular, it extends upon the models described by Escaff~\etal~\cite{escaff2020flocking} and Li~\etal~\cite{lowdimchiral}, and introduces triadic (\ie, higher-order) interactions in orientation dynamics originating via mechanisms described by Leon~\etal~\cite{leon2025collectivedynamicshigherordervicsek} and Marghoti~\etal~\cite{marghoti2026impactchanneldynamicshigherorder}, and investigates the order-disorder transitions occurring for different system parameter values and initial conditions. The remainder of this paper is organised as follows. Sec.~\ref{math} introduces the mathematical model under investigation, along with descriptions of various system parameters that govern the overall dynamics and of ensemble measures (\ie, order parameters) that quantify the macroscopic state of the system. Subsequently, Secs.~\ref{analytical} and \ref{simulations} outline the results obtained through analytical calculations under certain simplifying approximations and through detailed numerical simulations of the governing equations, respectively, with the latter section also highlighting the parameter regimes in which the numerical observations deviate from the theoretical predictions. The paper ends with Sec.~\ref{conc}, which provides a concise summary of the obtained results and highlights the scope for future work.

\section{Mathematical Model}
\label{math}

Assuming there are $N$ active particles in the system, occupying a flat square domain of side $L$ with periodic boundary conditions (PBCs), the spatial dynamics can be described by coupled ordinary differential equations of the form
\begin{align}
    \frac{\D x_i(t)}{\D t}=v_0 \cos\left(\psi_i\right),\quad
    \frac{\D y_i(t)}{\D t}=v_0 \sin\left(\psi_i\right),
\label{spatial-dynamics}
\end{align}
for all $i=1,2,\dots,N$. Here, $v_0$ is the common propulsion speed of all the particles, whereas $\psi_i(t)$ are the instantaneous heading angles. For the sake of simplicity, the governing equations for $\psi_i(t)$ are assumed to be a set of Kuramoto-like coupled ordinary differential equations having the form
\begin{align}
    \frac{\D \psi_i(t)}{\D t}=\omega_i&+\frac{\sigma_{\rm P}}{N}\sum_{\mathclap{j_1=1}}^N  F_{\rm P}\left(d_{ij_1}\right)\sin\left(\psi_{j_1}-\psi_i\right)\nonumber \\ &+\frac{\sigma_{\rm T}}{N^2}\sum_{\mathclap{j_1,j_2=1}}^N  F_{\rm T}\left(d_{ij_1},d_{ij_2}\right)\sin\left(2\psi_{j_2}-\psi_{j_1}-\psi_i\right)\nonumber\\  &+\sqrt{2D_{\psi}}\eta_i(t).
\label{phase-dynamics}
\end{align}
While the second and third terms in the RHS of Eq.~\eqref{phase-dynamics} capture the alignment interactions, fluctuations are introduced via the fourth term through a Gaussian white noise process $\eta_i(t)$ of strength $D_{\psi}$ that satisfies $\langle \eta_i(t)\rangle =0$ and $\langle \eta_i(t^{\prime})\eta_j(t)\rangle = \delta_{ij}\delta(t^{\prime}-t)$. 
Here, $d_{ij}$ is the instantaneous inter-particle distance matrix that implicitly couples spatial and orientational dynamics and is computed using the minimum-image convention for PBCs. On the other hand, $\omega_i$ are the constant uncoupled frequencies of self-rotation (\ie, chirality), and are sampled from a distribution $G(\omega)$, the exact form of which is described later. Furthermore, $F_{\rm P}$ and $F_{\rm T}$ capture the dependence of the pairwise and triadic coupling strengths on the distances between the particles participating in the interactions. Here, it is worth noting that while interactions of arbitrary order~\cite{biswas2026emergent} can be included in Eq.~\eqref{phase-dynamics}, the paper is restricted to triadic interactions for computational and analytical simplicity. More importantly, triadic interactions can also be shown to arise from fundamental system properties, with examples such as time-dependent pairwise interactions that incorporate local conformity interactions as well as those mediated by additional dynamical channels~\cite{leon2025collectivedynamicshigherordervicsek,marghoti2026impactchanneldynamicshigherorder}.

\begin{figure}[h]
    \centering
    \subfloat[\label{f1a}]{\includegraphics[width=0.335\linewidth]{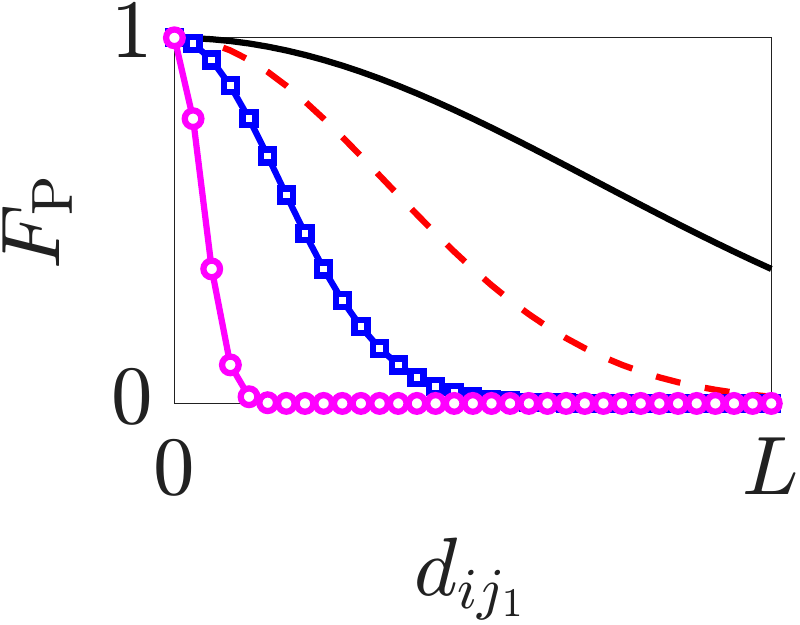}}
    \subfloat[$d_0=L/2$ (left), $L/4$ (right)\label{f1b}]{\includegraphics[width=0.335\linewidth]{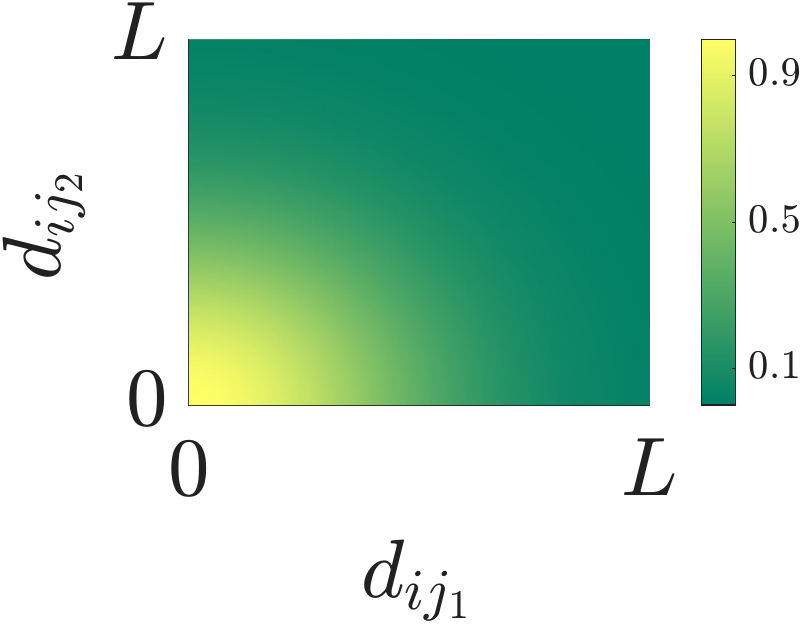}\includegraphics[width=0.335\linewidth]{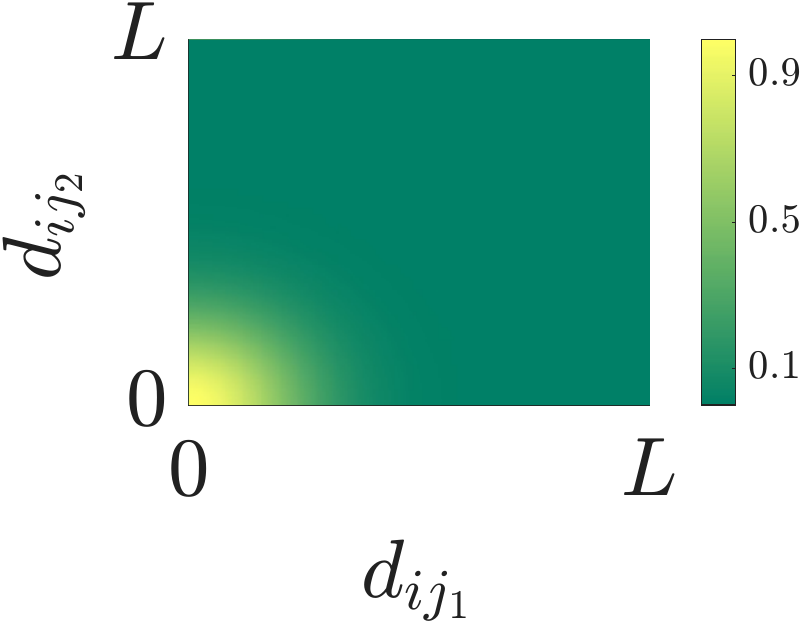}}

    \caption{
    (a): $F_{\rm P}$ as a function of interparticle distance $d_{ij_1}$. The curves indicate different values of $d_0=L$ (solid/black), $L/2$ (broken/red), $L/4$ (squares/blue), and $L/16$ (circles/magenta). (b): $F_{\rm T}$, represented as a color-map between $[0,1]$, as a function of $d_{ij_1}$ and $d_{ij_2}$, for different $d_0$. For all figures, the value of $L=32$.}
\end{figure}
Pairwise interactions are straightforward to implement, where, typically, the strength of any interaction is maximum when participating particles are very close and decreases rapidly with increasing inter-particle distance. This prompts the intuitive choice 
\begin{equation}
F_{\rm P}(d_{ij_1})=\exp\left(-d_{ij_1}^2/d^2_0\right), 
\label{dist-dependence}
\end{equation}
where $d_0$ captures the characteristic interaction length and is an important system parameter. The exponentially decaying form of the interaction kernel better approximates realistic interactions that taper off with distance (for a similar example, see the interaction function utilized by Bhaskar~\etal~\cite{x14x-yykw}), rather than the sharp top-hat function typically used in the literature~\cite{ginelli2016physics}.
On the other hand, for triadic interactions, it is \textit{assumed} for mathematical convenience that $F_{\rm T}\left(d_{ij_1},d_{ij_2}\right)= F_{\rm P}\left(d_{ij_1}\right)F_{\rm P}\left(d_{ij_2}\right)$, which satisfies the physical requirement $F_{\rm T}=1$ when the particles are coincident, and decreases smoothly with increasing distances; see Figs.~\ref{f1a} and \ref{f1b}. Furthermore, since $F_{\rm T}$ is expressed in terms of $F_{\rm P}$, in the rest of the paper, the subscripts of the interaction function $F$ are dropped for the sake of simpler notation. The subsequent section provides analytical insights into the system, with regard to its collective dynamics and the corresponding phase transitions.

\section{Analytical Insights}
\label{analytical}

In the large ensemble limit, the system can be described by a joint probability density $\mathscr{H}\left(\bm{r},\psi,t| \omega\right)$ which satisfies
\begin{equation}
    \int_{\mathscr{L}} \D^2 \bm{r} \int_{0}^{2\pi} \D\psi\ \mathscr{H}\left(\bm{r},\psi,t|\omega\right)=1,
\end{equation}
for every value of $\omega$, where the spatial integral is over the simulation box such that $\int_{\mathscr{L}} \D^2 \bm{r}=L^2$. Subsequently, the pairwise interactions in Eq.~\eqref{phase-dynamics} can be rewritten as $\sigma_{\rm P}\operatorname{Im}[Z^{(1)}_i e^{-\img \psi_i}]$ where $Z^{(1)}_i$ is the \textit{local} Kuramoto order parameter and $\operatorname{Im}(z)$ is the imaginary part of $z$. The former can be expressed as
\begin{subequations}
\begin{align}
    Z^{(1)}_i=&\frac{1}{N}\sum_{j=1}^N F\left(d_{ij}\right)e^{\img \psi_{j}}\\
    =&\int_{\mathscr{L}} \D^2\bm{r}_{j}\ F\left(d_{ij}\right) \int_{-\infty}^{\infty}\D\omega\ G(\omega) \int_{0}^{2\pi} \D\psi\ e^{\img \psi}\ \mathscr{H}.
\end{align}
\label{z1-local-expression}
\end{subequations}
Now, upon introducing the \textit{ansatz}
\begin{equation}
\mathscr{H}(\bm{r},\psi,t|\omega) = L^{-2}\ H(\psi,t|\omega),
\label{ansatz}
\end{equation}
which \textit{assumes} that the position and phase are statistically independent and the particles are distributed homogeneously in the simulation domain~\cite{lowdimchiral}, the expression in Eq.~\eqref{z1-local-expression} can be rewritten as
\begin{equation}
    Z^{(1)}_i=\frac{Z^{(1)}}{L^2}\int_{\mathscr{L}} \D^2\bm{r}_{j}\ F\left(d_{ij}\right)=\frac{Z^{(1)}}{L^2}I_i.
\label{Z1_i}
\end{equation}
Here, $Z^{(1)}$ is the \textit{global} Kuramoto order parameter given by
\begin{subequations}
\begin{align}
    Z^{(1)}=r^{(1)}e^{\img \Psi^{(1)}}&=\frac{1}{N}\sum_{j=1}^N e^{\img \psi_{j}}\\
           &=\int_{-\infty}^{\infty} \D\omega\ G(\omega)\int_{0}^{2\pi} \D\psi\ e^{\img \psi}\ H,
\end{align}
\end{subequations}
where $(r^{(1)},\Psi^{(1)})$ quantify the instantaneous synchronisation and average phase, respectively.
Similarly, the triadic interactions in Eq.~\eqref{phase-dynamics} can be recast as $\sigma_{\rm T}\operatorname{Im}[Z^{(2)}_i Z^{(1)*}_ie^{-\img\psi_i}]$, where $Z_i^{(2)}$ is an additional local Kuramoto-Daido order parameter~\cite{clusella2021irregular}. The latter can be expressed as
\begin{equation}
    Z^{(2)}_i=\frac{1}{N}\sum_{j=1}^N F(d_{ij})\ e^{2\img \psi_{j}}=\frac{Z^{(2)}}{L^2}I_i,
\label{Z2_i}
\end{equation}
where $Z^{(2)}=N^{-1}\sum_{j=1}^N e^{2\img\psi_{j}}$ is the corresponding global Kuramoto-Daido order parameter and $*$ indicates complex conjugation. Upon substituting, Eq.~\eqref{phase-dynamics} takes the form
\begin{align}
    \frac{\D \psi_i}{\D t}=\omega_i &+ \frac{\sigma_{\rm P}I_i}{L^2}\operatorname{Im}\left[Z^{(1)}e^{-\img\psi_i}\right]\nonumber\\ &+ \frac{\sigma_{\rm T} I^2_i}{L^4}\operatorname{Im}\left[Z^{(2)}\left(Z^{(1)}\right)^*e^{-\img\psi_i}\right]+\sqrt{2D_{\psi}}\eta_i(t),
\label{compact-form}
\end{align}
where $I_i=\pi d^2_0 E^2(L/2d_0)$ and $E(x)$ is the error function. Here, the integral $I_i$ is evaluated exactly, the details of which are presented in Appendix~\ref{critical-integral}. Thus, the system described by Eq.~\eqref{phase-dynamics} can be approximated as a triadic Kuramoto model with  \textit{effective} pairwise and triadic coupling strengths (say, $\sigma^e_{\rm P}$ and $\sigma^e_{\rm T}$) given by
\begin{equation}
    \sigma^{e}_{\rm P}=\frac{\pi d^2_0 E^2(L/2d_0)}{L^2}\sigma_{\rm P},\quad  
    \sigma^{e}_{\rm T}=\frac{\pi^2 d^4_0 E^4(L/2d_0)}{L^4}\sigma_{\rm T},
\label{effective-strength}
\end{equation}
respectively. Subsequently, these expressions can be used to estimate the critical phase transition point and quantify the steady-state order parameter dynamics, by using well-known results from the dynamics of the Kuramoto model.
\begin{figure}[h]
    \centering
    \subfloat[$\Omega=0$ (left), $\gamma=0.5$ (right)\label{fGplot1}]{\includegraphics[width=0.33\linewidth]{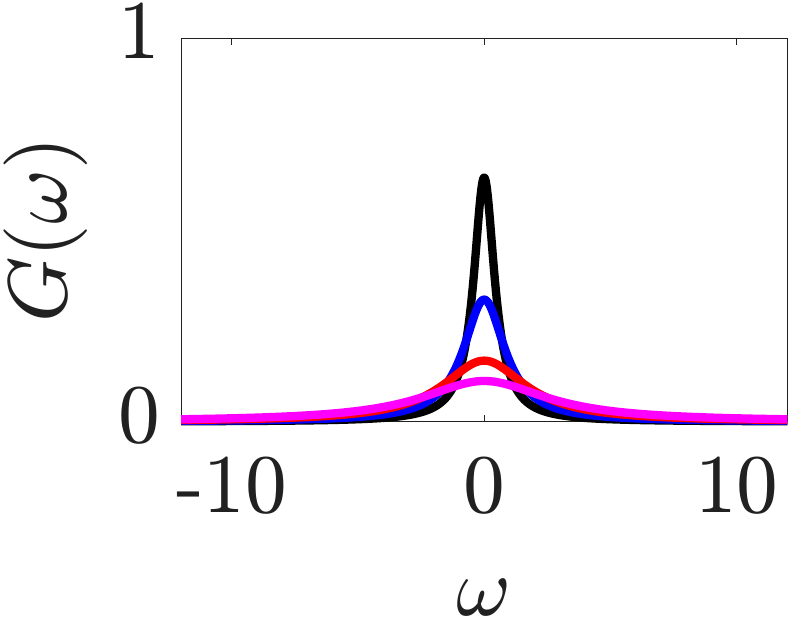}\includegraphics[width=0.33\linewidth]{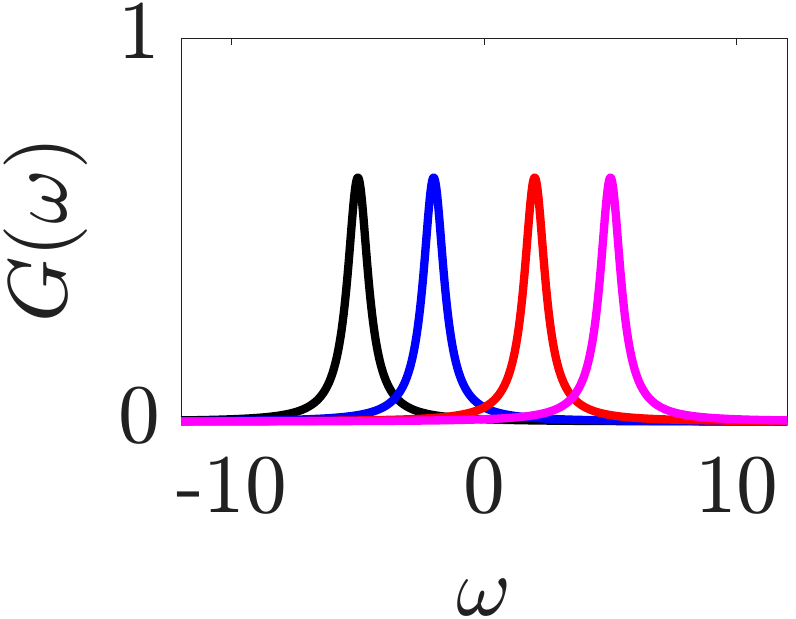}}
    \subfloat[$(\gamma,D_{\psi})=(2.0,1.0)$\label{sigma_pc_limit}]{\includegraphics[width=0.32\linewidth]{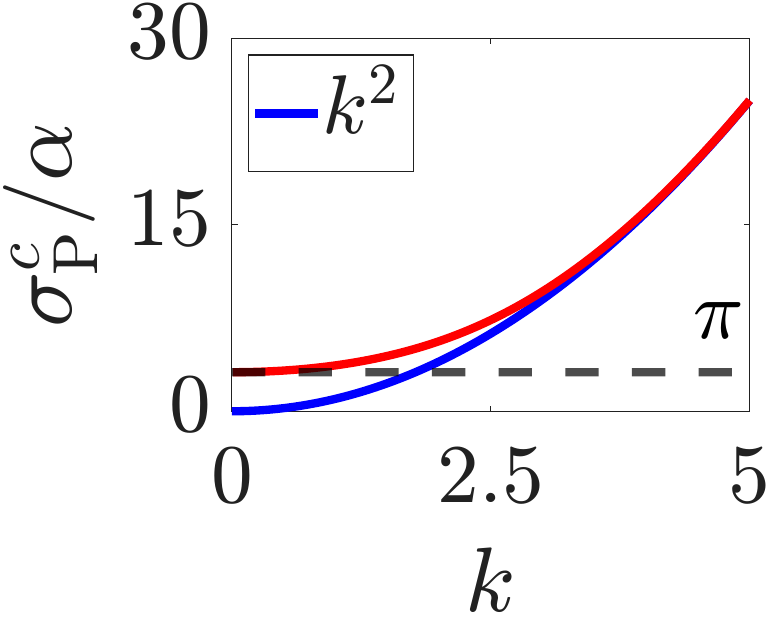}}

    \subfloat[$\gamma=0.01$, $0.1$, and $1.0$ (left to right)\label{sigma_pc_vs_kD}]{\includegraphics[width=0.33\linewidth]{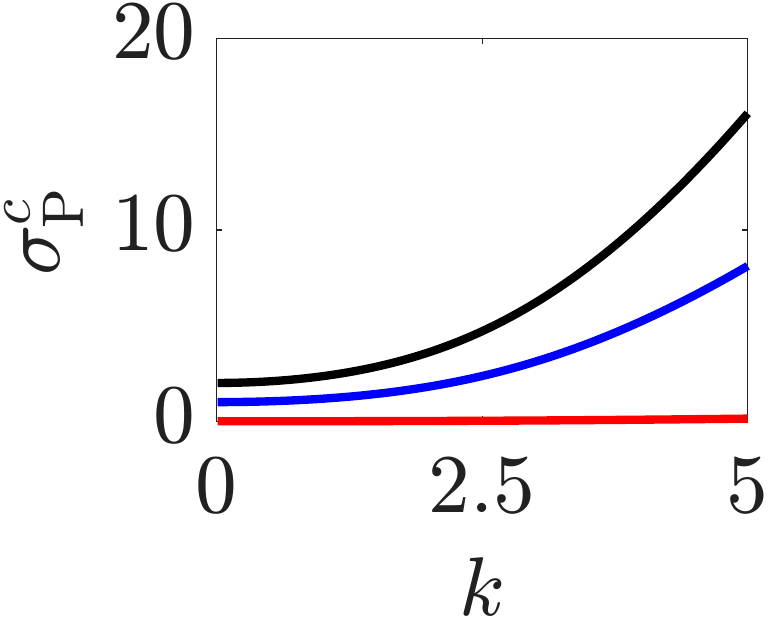}\hfill\includegraphics[width=0.33\linewidth]{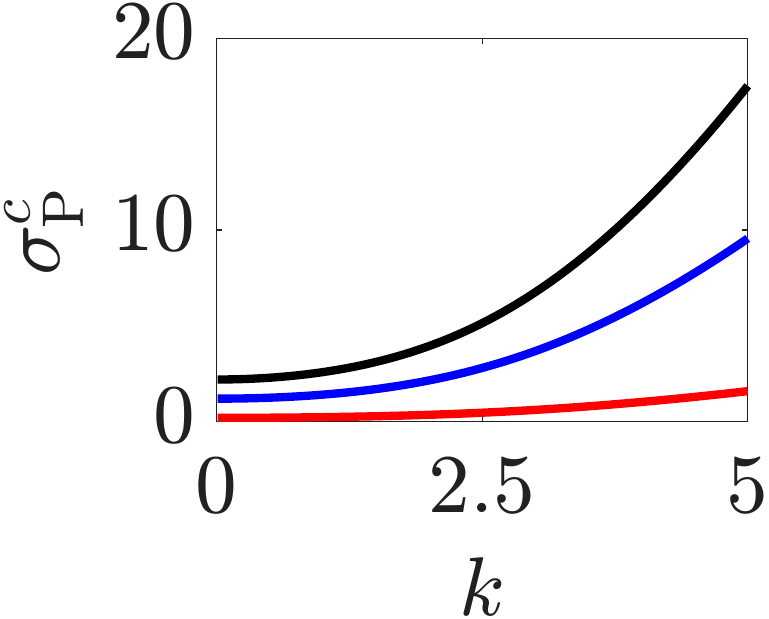}\hfill\includegraphics[width=0.33\linewidth]{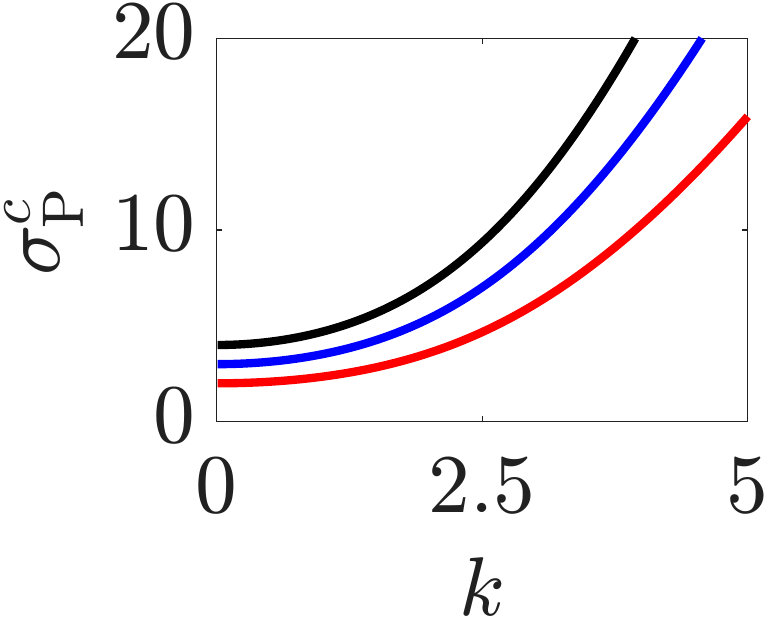}}

    \caption{(a): $G$ as a function of $\omega$ for $\Omega=0$ and $\gamma=0.5$, respectively. In the left sub-figure, $\gamma=0.5$ (black), $1.0$ (blue), $2.0$ (red), and $3.0$ (magenta), whereas in the right sub-figure, $\Omega=-5.0$ (black), $-2.0$ (blue), $2.0$ (red), and $5.0$ (magenta); (b): $\sigma^c_{\rm P}/\alpha$ (from Eq.~\eqref{limit-cases}) as a function of $k$ (in red) for constant $(\gamma,D_{\psi})$, where the horizontal line and blue curve denote $k\rightarrow 0$ and $k\gg 0$, respectively (from Eq.~\eqref{limit-cases}); (c): $\sigma^c_{\rm P}$ as a function of $k$ for different values of $\gamma$, where the colors indicate $D_{\psi}=0.0$ (red), $0.5$ (blue), and $1.0$ (black).}
\end{figure}

Upon assuming that the characteristic interaction range is parameterized in terms of the box size by $k$ such that $d_0=L/k$ and $G(\omega)$ is a Lorentzian distribution of mean $\Omega\in\mathbb{R}$ and scale factor $\gamma\in \mathbb{R}^+$ (see Fig.~\ref{fGplot1}), the critical phase transition point $\sigma_{\rm P}=\sigma^{c}_{\rm P}$ at which the asynchronous state $r^{(1)}(t)=0$ loses stability can be obtained as
\begin{align}
    \sigma^{c}_{\rm P}&=\frac{2L^2 (\gamma + D_{\psi})}{\pi d^2_0 E^2(L/2d_0)}= \frac{\alpha k^2}{E^{2}(k/2)},
\label{forward-trans-point}
\intertext{where $\alpha=2\pi^{-1}(\gamma+D_{\psi})$ for the sake of compact notation. In limiting cases, the value of $\sigma^c_{\rm P}$ varies as}
    \frac{\sigma^{c}_{\rm P}}{\alpha}&=
    \begin{cases}
        \pi ,\ &k\rightarrow 0,\\
        k^2,\ &k\gg 0,
    \end{cases}
\label{limit-cases}
\intertext{
where the traditional Kuramoto phase transition point $\sigma^c_{\rm P}=\pi\alpha=2(\gamma+D_{\psi})$ is recovered for long-range interactions (\ie, $k\rightarrow 0$). The analysis leading to Eq.~\eqref{forward-trans-point} also indicates that}
    D^c_{\psi}&=\frac{\Delta}{2}\ \beta(k),
\label{critical_noise}
\end{align}
where $D^c_{\psi}$ is the critical noise level for a system with $\sigma_{\rm P}=\sigma^c_{\rm P}(D_{\psi}=0)+\Delta$. Here, $\Delta>0$ controls how deep in the synchronized regime the system is initialized, and the function $\beta(k)=\pi k^{-2} E^2(k/2)$ captures the dependence on $k$.

The theoretically obtained variation of $\sigma^c_{\rm P}$ is visualized in Figs.~\ref{sigma_pc_limit} and \ref{sigma_pc_vs_kD}, where, the former highlights the limiting cases $k\rightarrow 0$ and  $k\gg 1$ for constant $(\gamma,D_{\psi})$, whereas the latter highlights its dependence across different combinations of $(\gamma,D_{\psi})$. Taken together, these results indicate that either reducing interaction range (\ie, increasing $k$) or increasing system disorder (\ie, by increasing $D_{\psi}$, $\gamma$, or both) necessitates stronger interactions to induce synchronisation, thereby increasing $\sigma^c_{\rm P}$. The expression in Eq.~\eqref{forward-trans-point} also indicates that the value of $\sigma^{c}_{\rm P}$ is independent of $\sigma_{\rm T}$. Heuristically, this can be attributed to higher-order interactions that contribute only nonlinear terms to the order-parameter dynamics, thereby affecting the value of the steady-state order parameter post-transition. It also affects the nature of the transition (\ie, continuous or explosive, the latter accompanied by the formation of a bistable zone and a hysteresis loop) and the backward transition. This is apparent from the equation governing the order parameter dynamics~\cite{ott2008low,ott2009long,biswas2024symmetry,biswas2026emergent,sinha2023statistics,biswas2026role}, which takes the form 
\begin{align}
&\frac{\D r}{\D t}=-(\gamma+D_{\psi}) r + \sigma_{\rm P}\ \beta(k)\frac{r(1-r^2)}{2}\Big[1+\sigma_{\rm R}\ \beta(k)r^2\Big],
\label{r_time}\\
\intertext{where $r(t)\equiv r^{(1)}(t)$. This equation admits a steady-state solution $r_{\rm ss}=+\sqrt{x}\in[0,1]$ that satisfies}
&\frac{2(\gamma+D_{\psi})}{\left(1-x\right)}=\sigma_{\rm P}\ \beta(k)\ \Big[1+x\ \sigma_{\rm R}\ \beta(k) \Big],
\label{rss_eqn}
\end{align}
which is stable $\forall\ 2(\gamma+D_{\psi}) > \sigma_{\rm R} \sigma_{\rm P}\ \beta^2(k) \left(1-x\right)^2$, where $\sigma_{\rm R}=\sigma_{\rm T}/\sigma_{\rm P}$ is the relative coupling strength.  Here, it is important to note that the above two equations are exact only for $D_{\psi}=0$, and an approximation 
that deteriorates for $D_{\psi}>0$.
\begin{figure}[h]
    \centering

    \subfloat[$D_{\psi}=0.0$\label{rss_plot_theo_a}]{\includegraphics[width=0.34\linewidth]{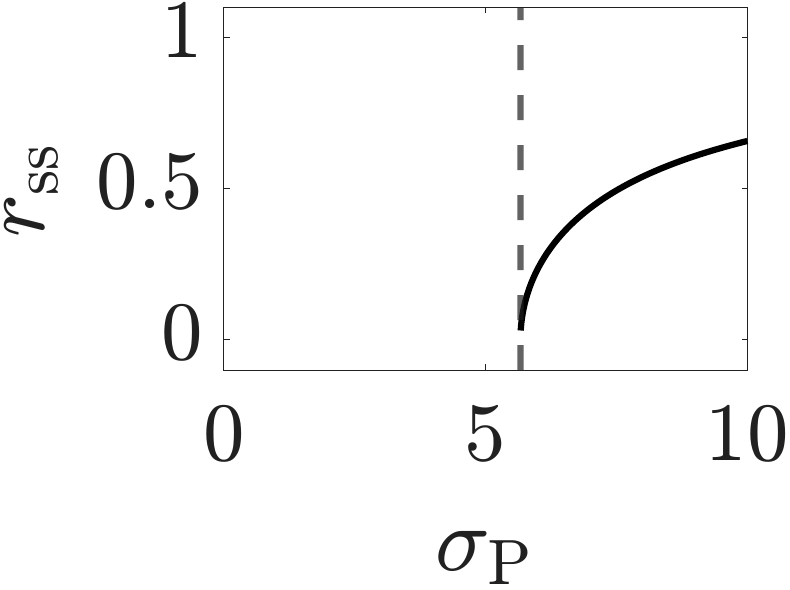}\includegraphics[width=0.34\linewidth]{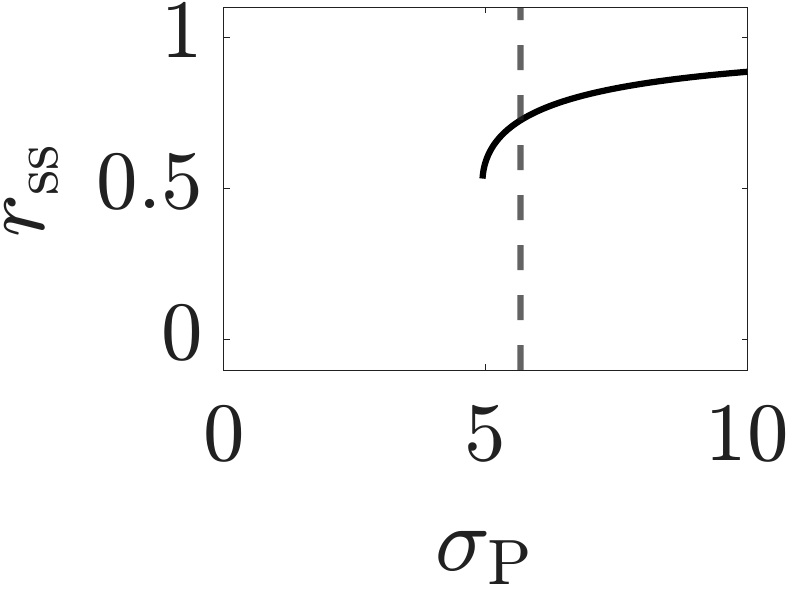}\includegraphics[width=0.34\linewidth]{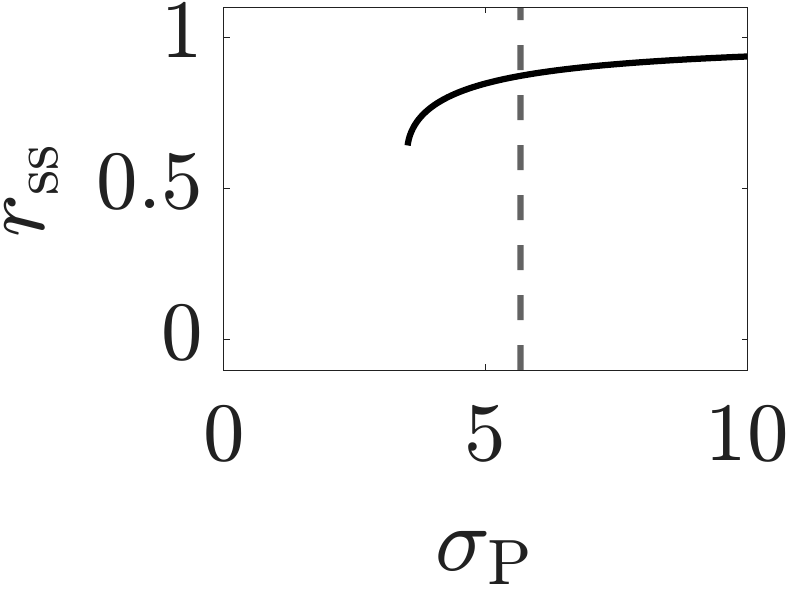}}
    
    \subfloat[$D_{\psi}=1.0$\label{rss_plot_theo_b}]
    {\includegraphics[width=0.34\linewidth]{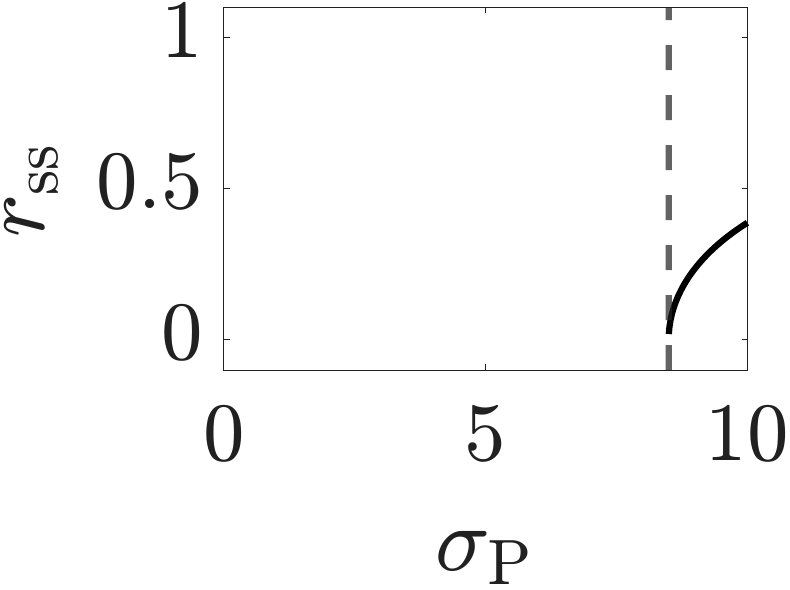}\includegraphics[width=0.34\linewidth]{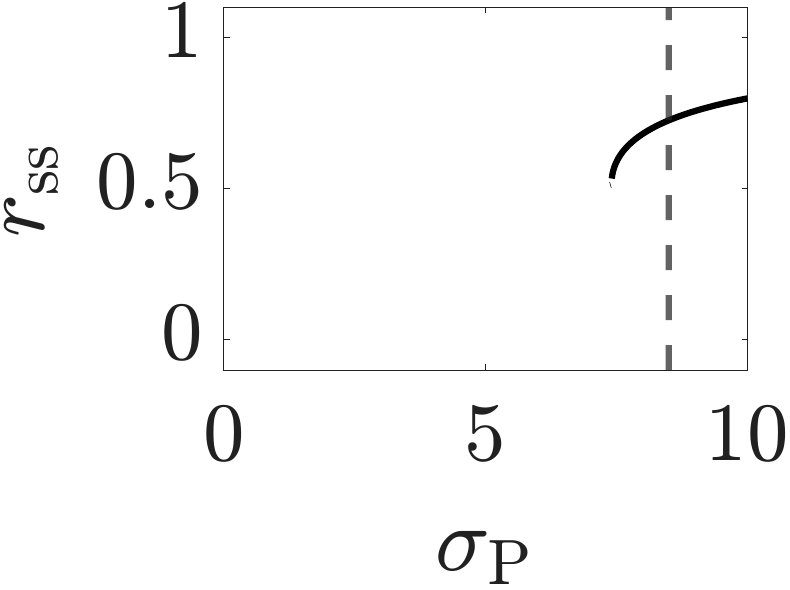}\includegraphics[width=0.34\linewidth]{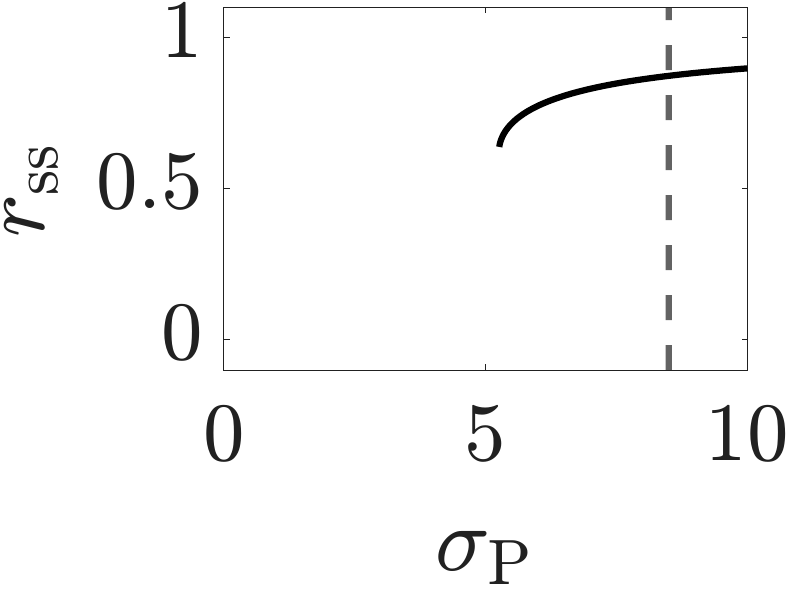}}

    \subfloat[$\sigma_{\rm R}=6.0$, $D_{\psi}=1.0$\label{rt_plot_theo}]{\includegraphics[width=0.4\linewidth]{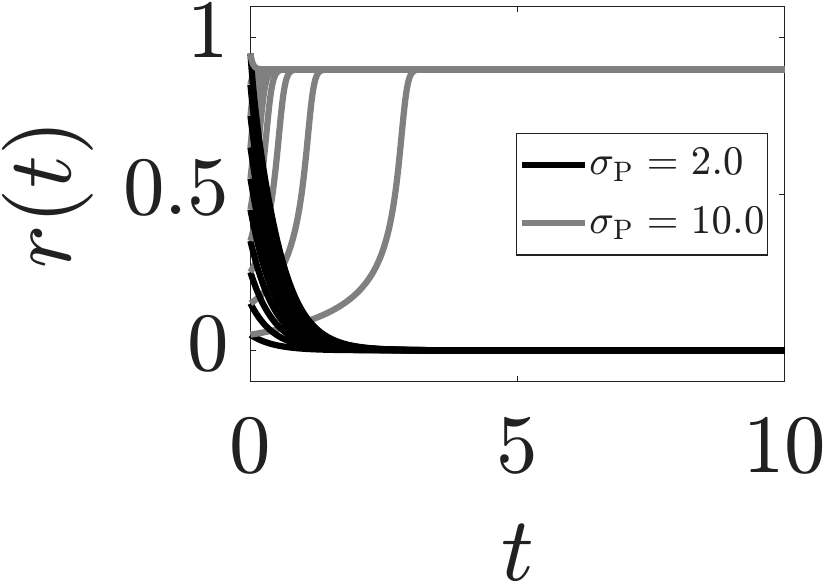}\hfill\includegraphics[width=0.4\linewidth]{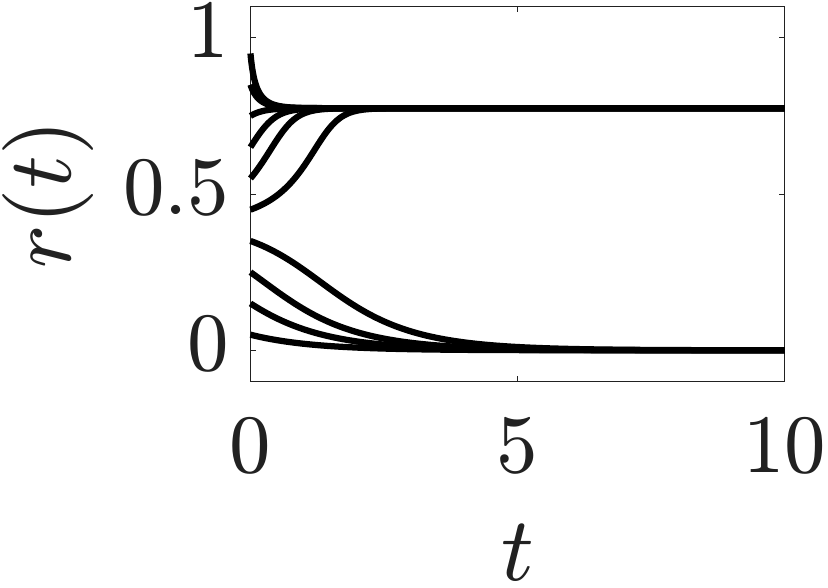}}
    
    \caption{(a)-(b): $r_{\rm ss}$ (from Eq.~\eqref{rss_eqn}) as a function of $\sigma_{\rm P}$ for different values of $D_{\psi}=0$ and $1.0$, respectively, whereas $\sigma_{\rm R}=0.0$ (left), $3.0$ (center), and $6.0$ (right). Here, the vertical line denotes $\sigma^c_{\rm P}$; (c): Realizations of $r(t)$ as a function of $t$ (from Eq.~\eqref{r_time}), for constant $\sigma_{\rm R}$ and different values of $\sigma_{\rm P}$, whereas each individual curve corresponds to different values of $r(0)\in(0,1)$. In the left panel, the black and grey curves correspond to $\sigma_{\rm P}=2.0$ and $10.0$, respectively, whereas the black curves in the right panel correspond to $\sigma_{\rm P}=6.0$. Across all figures, the values of $(\gamma,k)=(2.0,1.5)$.}
    \label{theo-phase-transitions}
\end{figure}

The variation of $r_{\rm ss}$ as function of $\sigma_{\rm P}$, obtained from Eq.~\eqref{rss_eqn}, is demonstrated in Figs.~\ref{rss_plot_theo_a} and \ref{rss_plot_theo_b} for $D_{\psi}=0.0$ and $1.0$, respectively, for different values of $\sigma_{\rm R}$. The curves indicate two qualitatively different types of phase transitions: for $\sigma_{\rm R}=0$, the transition is continuous; for $\sigma_{\rm R}=3.0$ and $6.0$, the transition is explosive, along with the formation of a bistable hysteresis zone, the size of which increases with $\sigma_{\rm R}$. The effect of bistability is further demonstrated in Fig.~\ref{rt_plot_theo} through the temporal variation of $r(t)$, obtained from Eq.~\eqref{r_time}, for three qualitatively different values of $\sigma_{\rm P}$. For $\sigma_{\rm P}=2.0$ and $10.0$, the system is monostable and converges to an identical steady state irrespective of $r(0)$; in contrast, for $\sigma_{\rm P}=6.0$, the system is bistable and converges to non-identical steady states depending on $r(0)$. The next section uses direct numerical simulations of the governing ordinary differential equations to corroborate the analytics, in addition to delineating the regime of validity of the theoretical predictions.

\section{Numerical results}
\label{simulations}

\begin{figure}[t]
    \centering
        
    \subfloat[$D_{\psi}=0.0$\label{pt_common_a}]{\includegraphics[width=0.33\linewidth]{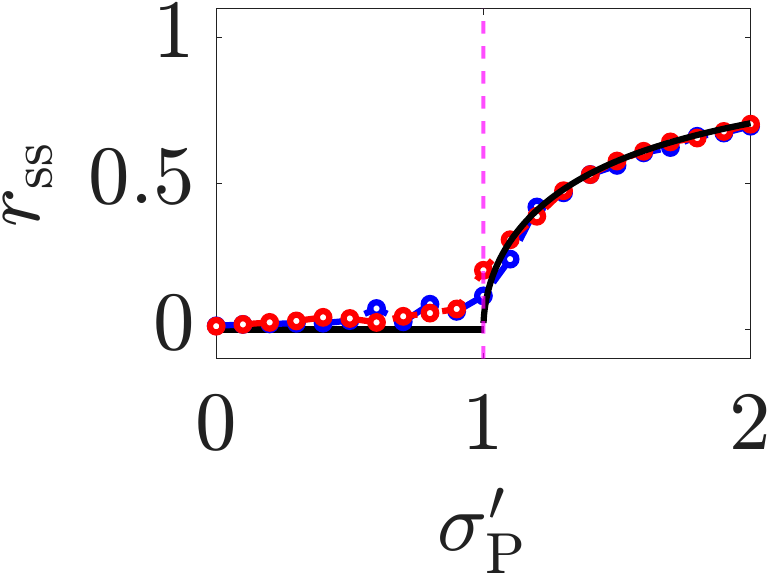}\includegraphics[width=0.33\linewidth]{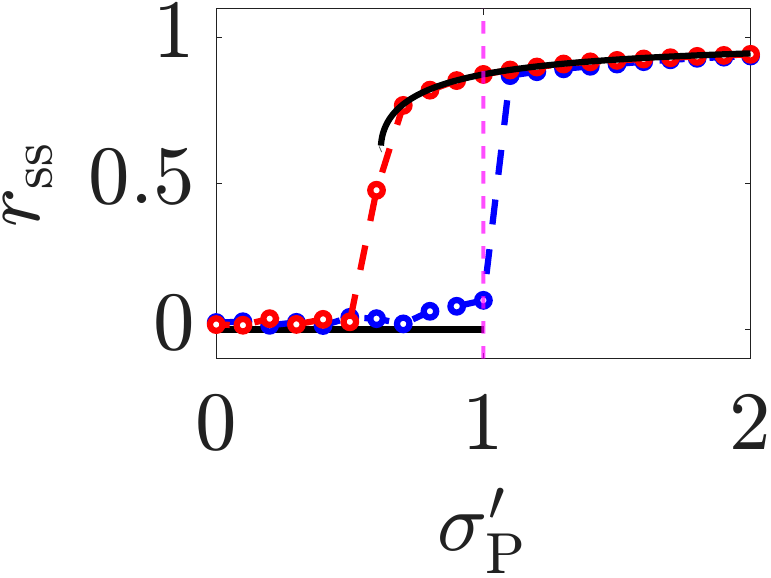}\includegraphics[width=0.33\linewidth]{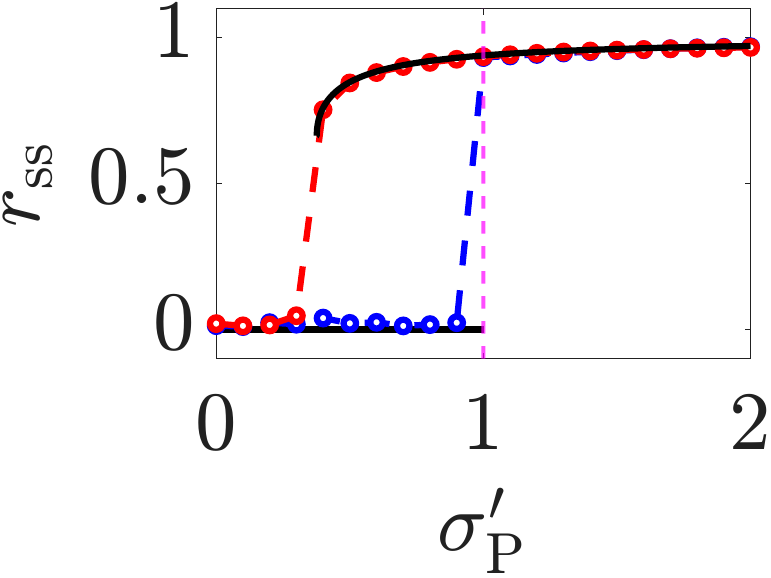}}
    
    \subfloat[$D_{\psi}=1.0$\label{pt_common_b}]{\includegraphics[width=0.33\linewidth]{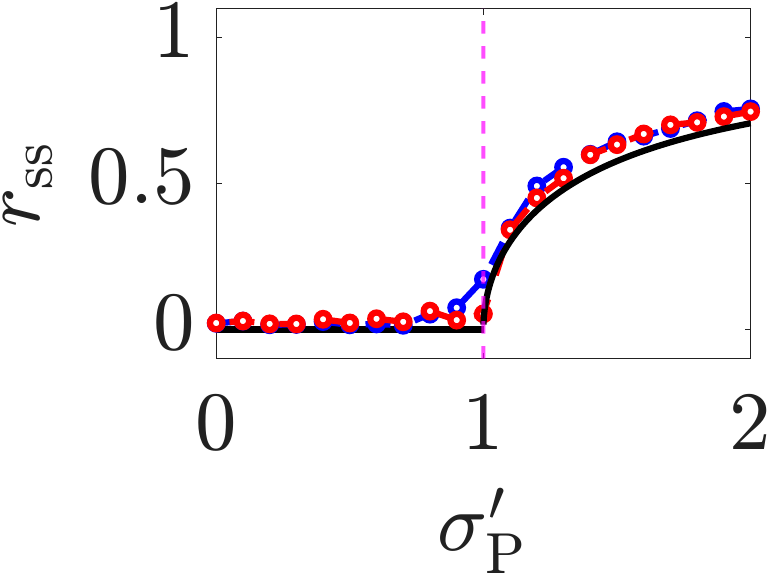}\includegraphics[width=0.33\linewidth]{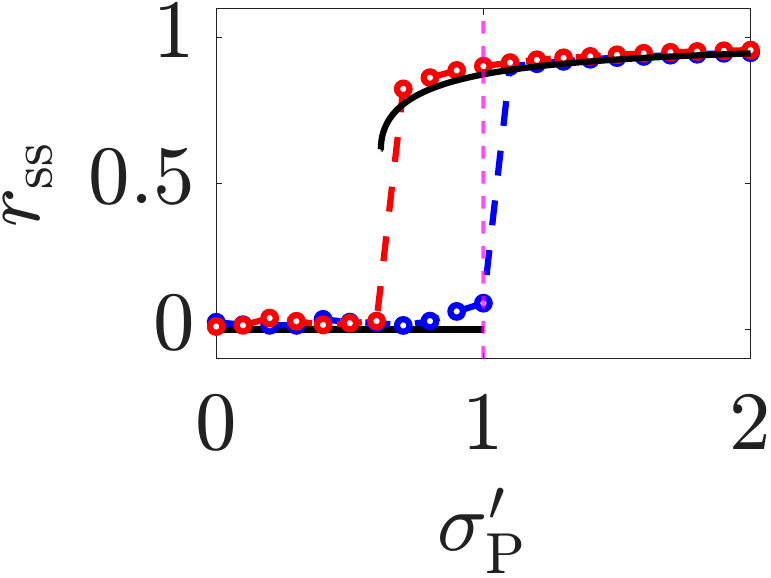}\includegraphics[width=0.33\linewidth]{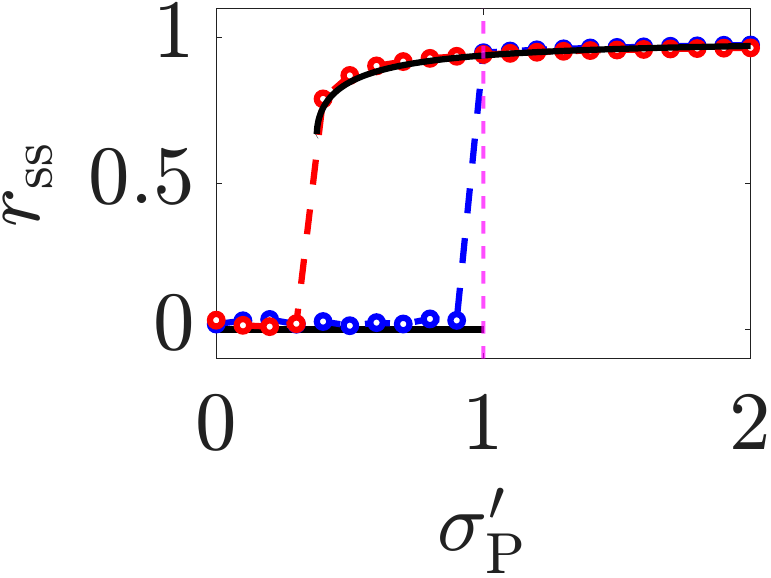}}

    \subfloat[$D_{\psi}=5.0$\label{pt_common_c}]{\includegraphics[width=0.33\linewidth]{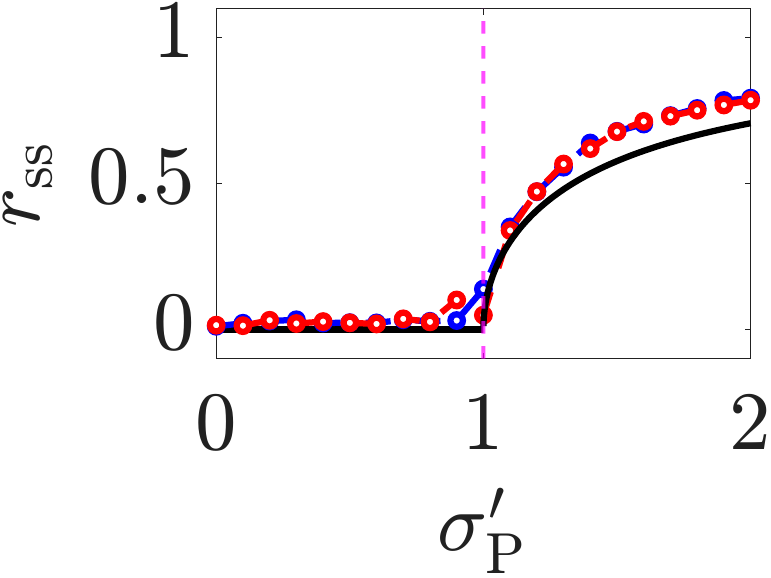}\includegraphics[width=0.33\linewidth]{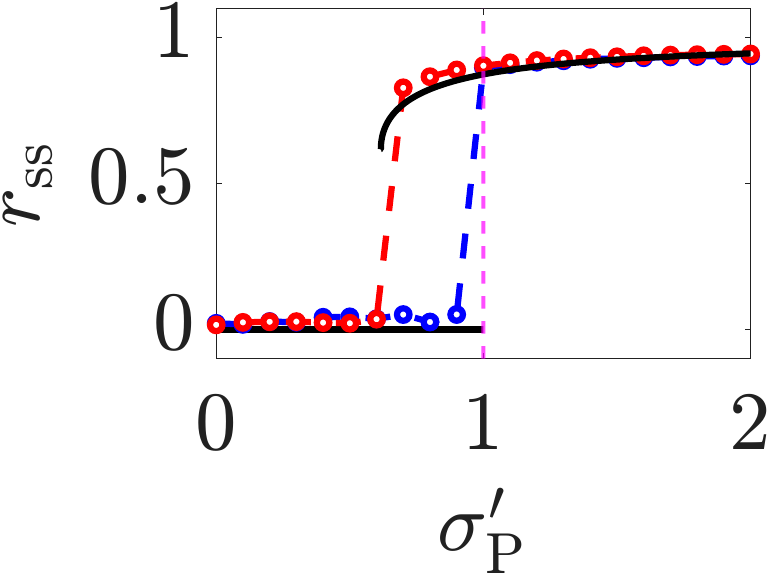}\includegraphics[width=0.33\linewidth]{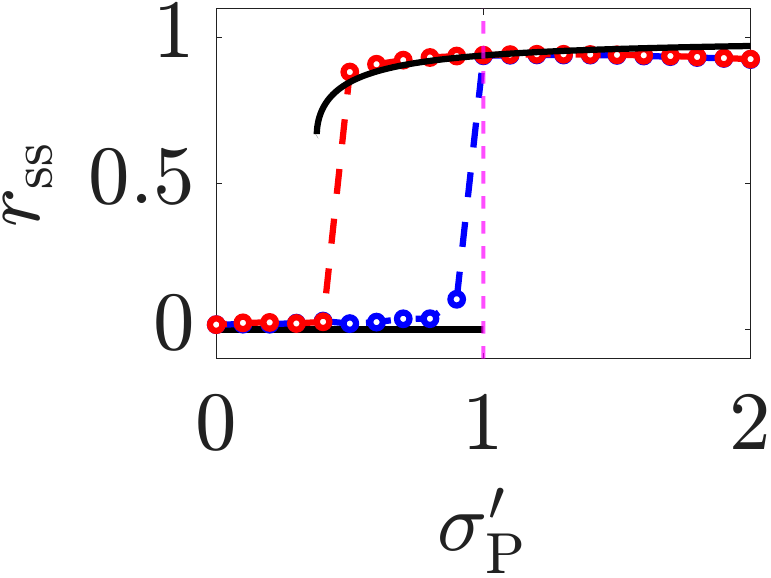}}

    \caption{(a)-(c): $r_{\rm ss}$ as a function of $\sigma^{\prime}_{\rm P}$ for $D_{\psi}=0.0$ (top row), $1.0$ (middle row), and $5.0$ (bottom row), respectively. Here, the blue and red curves denote the forward and backward parameter sweeps, respectively, the vertical magenta line marks the theoretical estimate of the forward phase-transition point (from Eq.~\eqref{forward-trans-point}), and the solid black curve denotes the theoretical estimate of $r_{\rm ss}$ (from Eq.~\eqref{rss_eqn}). The sub-figures correspond to $\sigma_{\rm R}=0$ (left), $5$ (center), and $10$ (right), whereas $(\gamma,\Omega,k)=(1.0,0.0,1.0)$ and $(T_{\rm R},T_{\rm C})=(10^3,10^2)$ for all.}
    \label{phase-transition-common}
\end{figure}

The Eqs.~\eqref{spatial-dynamics} and \eqref{phase-dynamics} are simulated in MATLAB, using a parallelized Euler-Maruyama method, with a time step of $\delta t=10^{-2}$. Unless mentioned otherwise, the common propulsion speed $v_0=1$, ensemble size $N=2048$, and box size $L=32$  (thus, density $\rho=N/L^2=2.0$) are used in the rest of the article, the rationale for which is outlined in Appendix \ref{sys-size}.  Additionally, the particles are initially distributed over the entire space (an assumption relaxed later in the article), and the interaction strengths are considered to be non-negative. For the sake of robustness, the steady-state value $r_{\rm ss}$ of the synchronisation order parameter $r(t)$ is computed by averaging over $T_{\rm C}$ steps, after discarding initial $T_{\rm R}$ steps to minimize the effect of transients.

To validate Eq.~\eqref{forward-trans-point}, Fig.~\ref{phase-transition-common} plots the variation of $r_{\rm ss}$, obtained through the direct numerical simulations of the governing equations, as a function of $\sigma^{\prime}_{\rm P}=\sigma_{\rm P}/\sigma^c_{\rm P}\in[0,2]$ for deterministic (\ie, $D_{\psi}=0$ in Fig.~\ref{pt_common_a}) and stochastic cases (\ie, $D_{\psi}=1.0$ and $5.0$ in Figs.~\ref{pt_common_b} and \ref{pt_common_c}), respectively\footnote{The numerically computed phase transitions are obtained as a function of $\sigma^{\prime}_{\rm P}=\sigma_{\rm P}/\sigma^c_{\rm P}$ (or, $D^{\prime}_{\psi}=D_{\psi}/D^c_{\psi}$), so that the transition always occurs at $\sigma^{\prime}_{\rm P}=1.0$ (or, $D^{\prime}_{\psi}=1.0$), irrespective of other system parameters. Furthermore, the parameter sweeps can be implemented by either (i) initializing the system with $r(0)\approx 0$ (forward) and $r(0)\approx 1$ (backward), respectively; or (ii) using the final system state as the initial condition for the subsequent simulation for increasing (forward) and decreasing (backward) control parameter values (\ie, $\sigma^{\prime}_{\rm P}$ or $D^{\prime}_{\psi}$). While these will produce practically identical phase transition diagrams for this system, the former is used in this paper since it allows parallelization across control parameter values.}. The key observation is the excellent agreement between the predicted (vertical line) and numerically obtained phase-transition points across all parameter combinations, as well as the value of the steady-state order parameter. The curves also indicate that for $\sigma_{\rm R}=0$, the transition is smooth, with backward (red) and forward (blue) parameter sweeps coinciding, regardless of whether the system is deterministic or stochastic. This is in contrast to the case for $\sigma_{\rm R}\neq 0$, where the transition can become explosive and manifests as a sudden change in $r_{\rm ss}$ at both the forward and backward phase transition points. Furthermore, the backward phase transition occurs at a different $\sigma^{\prime}_{\rm P}$, indicating the emergence of bistability, with the area of the hysteresis loop increasing with increasing $\sigma_{\rm R}$.
\begin{figure}[h]
    \centering

    \subfloat[$\Delta=2.0$; $\sigma_{\rm R}=0.0$ (left), $2.5$ (center), and $5.0$ (right).\label{Delta_2}]{\includegraphics[width=0.33\linewidth]{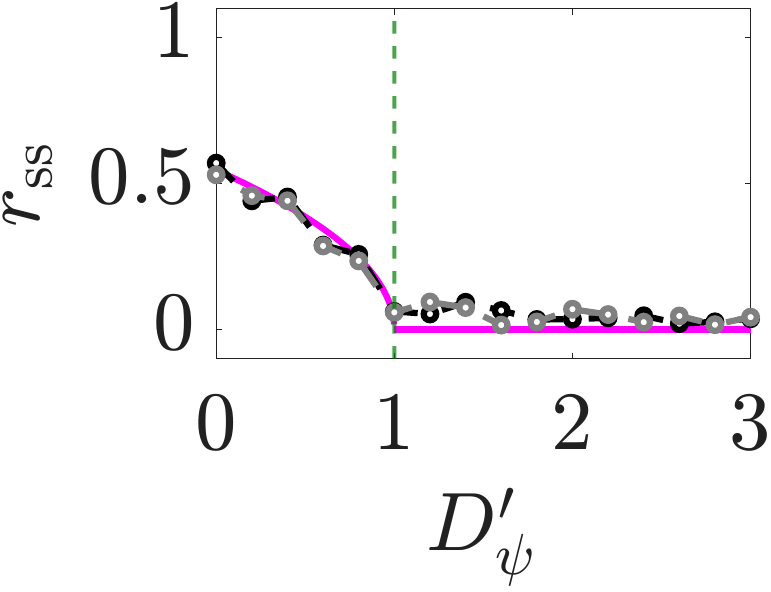}\includegraphics[width=0.33\linewidth]{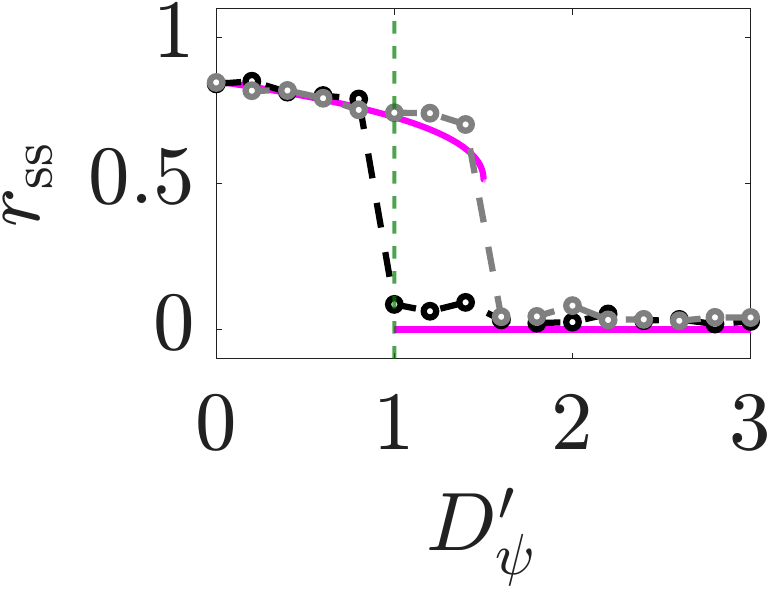}\includegraphics[width=0.33\linewidth]{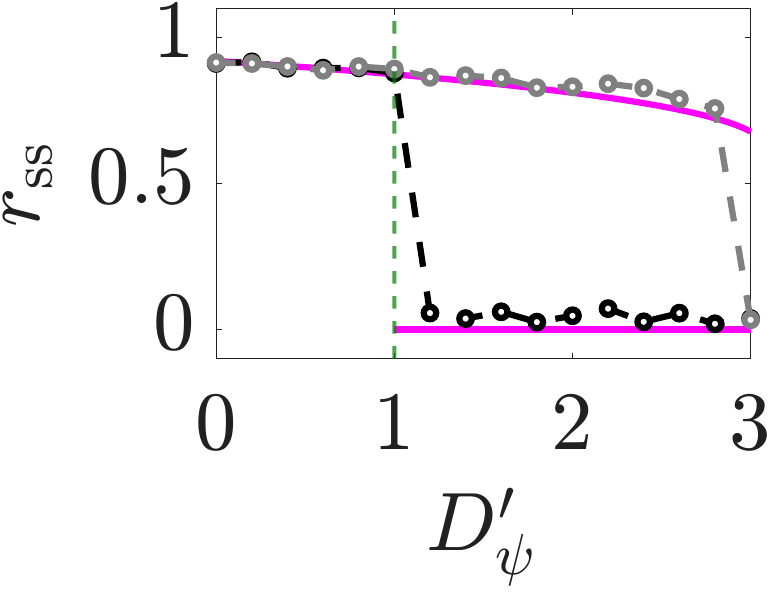}}
    
    \subfloat[$\Delta=5.0$; $\sigma_{\rm R}=0.0$ (left), $3.0$ (center), and $6.0$ (right).\label{Delta_5}]{\includegraphics[width=0.33\linewidth]{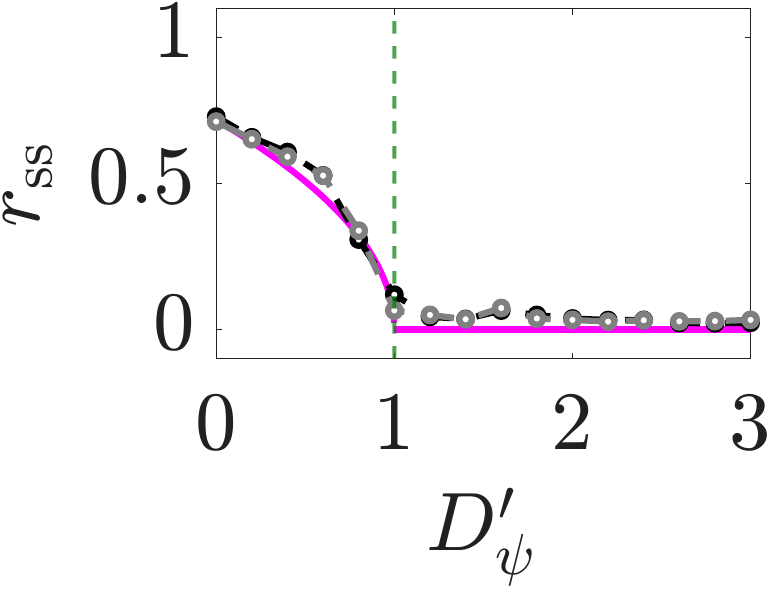}\includegraphics[width=0.33\linewidth]{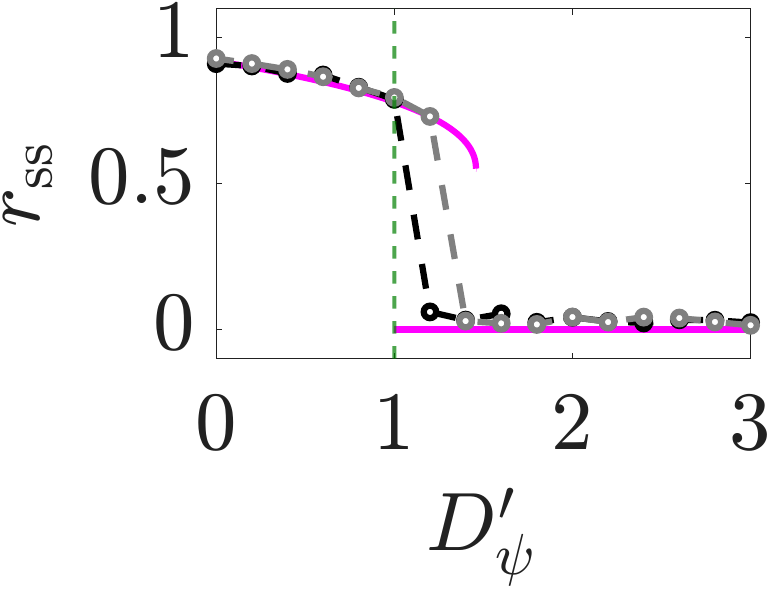}\includegraphics[width=0.33\linewidth]{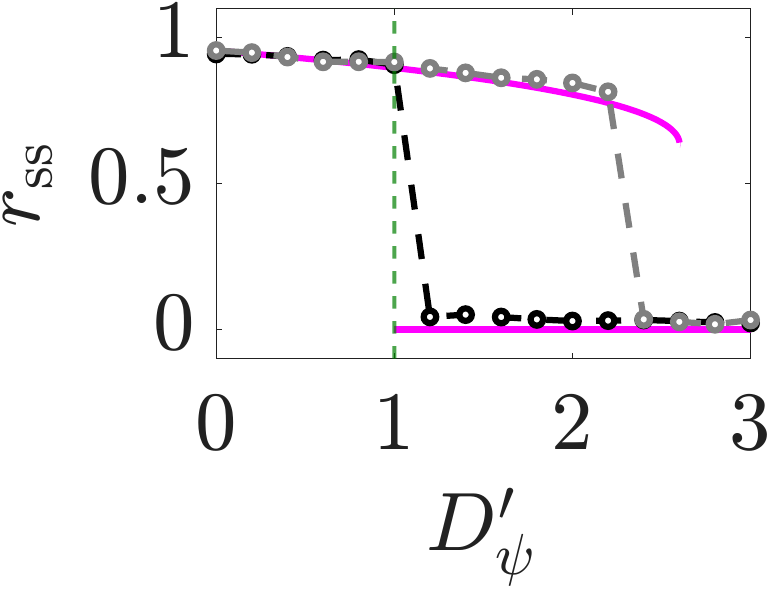}}

    \subfloat[$\Delta=10.0$; $\sigma_{\rm R}=0$ (left), $5.0$ (center), and $10.0$ (right).\label{Delta_10}]{\includegraphics[width=0.33\linewidth]{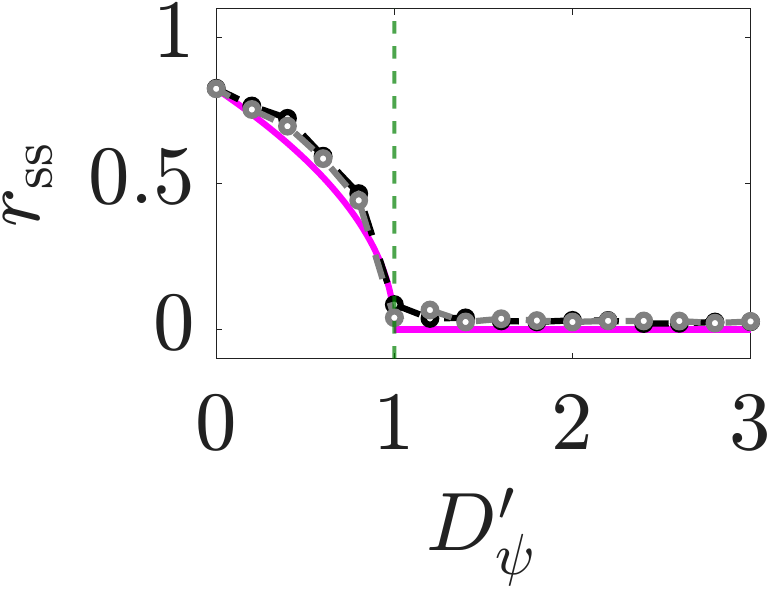}\includegraphics[width=0.33\linewidth]{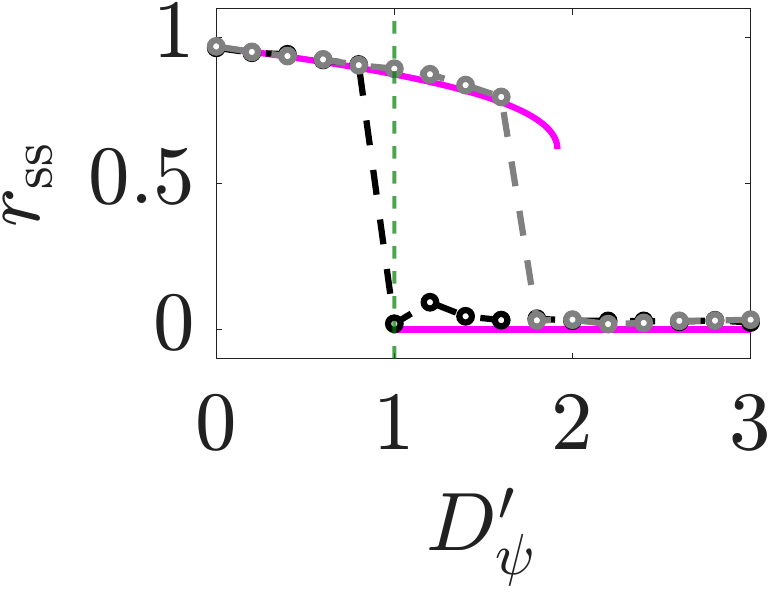}\includegraphics[width=0.33\linewidth]{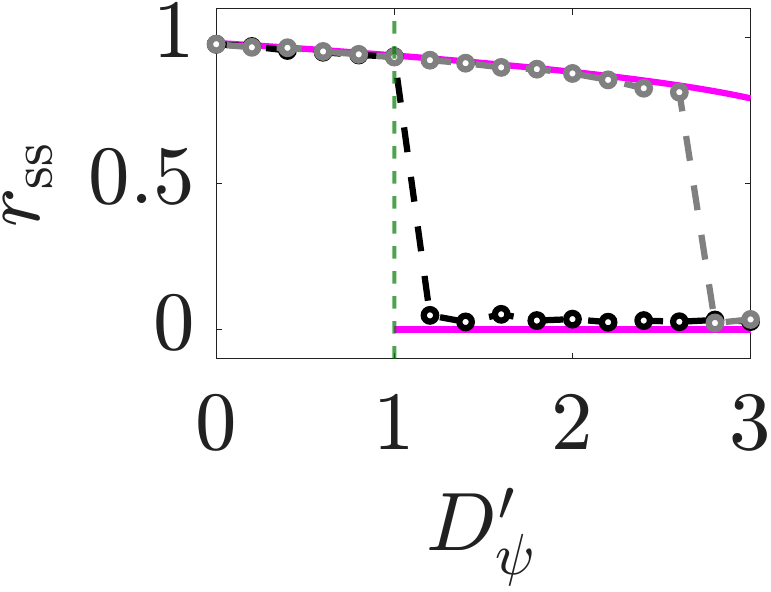}}

\caption{(a)-(c): $r_{\rm ss}$ as a function of $D^{\prime}_{\psi}$ for different values of $\sigma_{\rm R}$ and $\Delta$. Here, the black and grey curves denote the forward and backward parameter sweeps, respectively, the vertical line corresponds to the theoretical estimate of the forward phase-transition point (from Eq.~\eqref{critical_noise}), and the magenta curve denotes the theoretical estimate of $r_{\rm ss}$ (from Eq.~\eqref{rss_eqn}). For all sub-figures, $(\gamma,\Omega,k)$ and $(T_{\rm R},T_{\rm C})$ are identical to Fig.~\ref{phase-transition-common}.}
\label{noise-plots-computed}
\end{figure}

Subsequently, to validate Eq.~\eqref{critical_noise}, phase transition diagrams (similar to Fig.~\ref{phase-transition-common}) are obtained in Fig.~\ref{noise-plots-computed}, however, as a function of $D^{\prime}_{\psi}=D_{\psi}/D^c_{\psi}\in[0,3]$, for $\Delta=2.0$ (in Fig.~\ref{Delta_2}), $\Delta=5.0$ (in Fig.~\ref{Delta_5}), and $\Delta=10.0$ (in Fig.~\ref{Delta_10}), with other parameters remaining identical. As is expected, the curves indicate that increasing noise intensity drives the system from order to disorder. Furthermore, for $\sigma_{\rm R}=0$ and $\sigma_{\rm R}\neq 0$, the nature of the transition is continuous and explosive, respectively, with the emergence of a bistable zone in the latter.
\begin{figure}[b]
    \centering

    \subfloat[$(\sigma_{\rm R},D_{\psi})=(10.0,1.0)$; $\sigma^{\prime}_{\rm P}=0.1$ (left), $0.9$ (center), and $1.9$ (right).\label{bistab_ts_a}]{\includegraphics[width=0.33\linewidth]{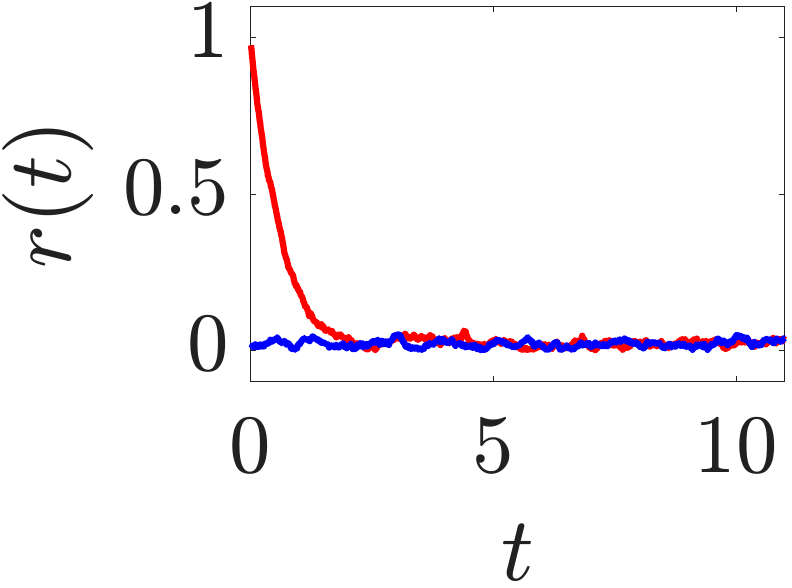}\includegraphics[width=0.33\linewidth]{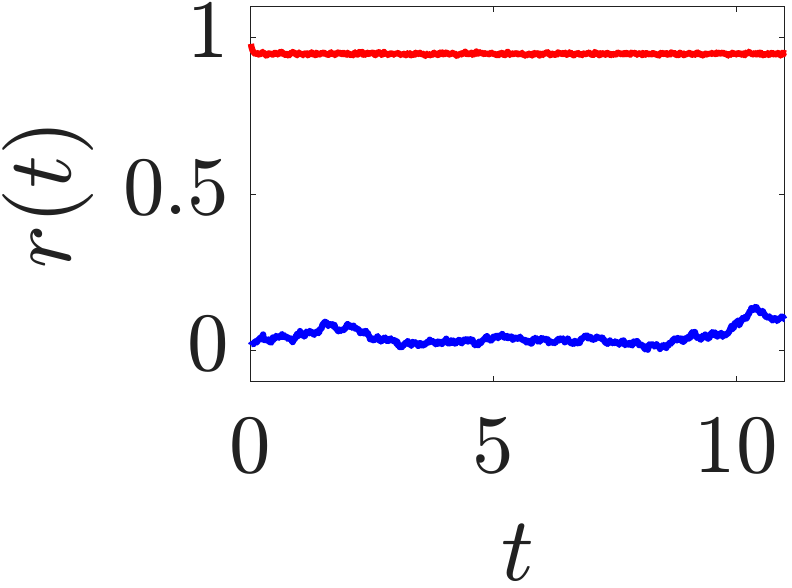}\includegraphics[width=0.33\linewidth]{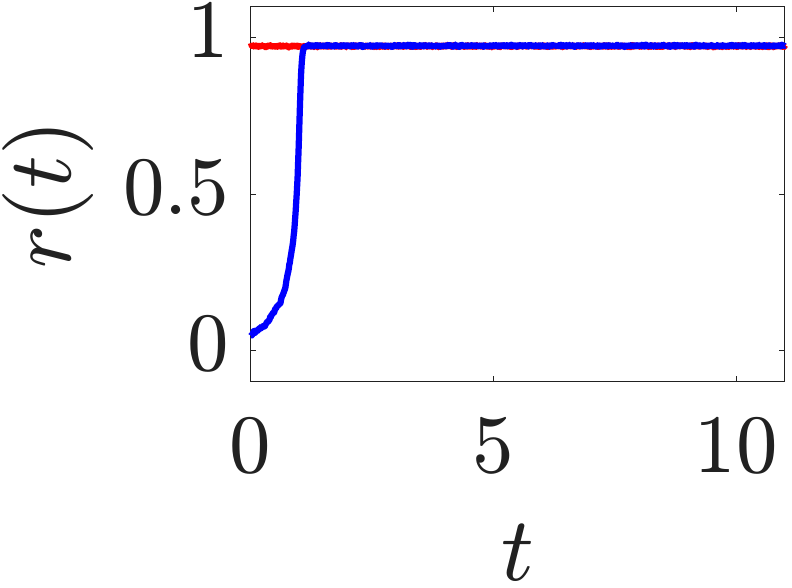}}

    \subfloat[$(\sigma_{\rm R},\Delta)=(5.0,10.0)$; $D^{\prime}_{\psi}=0.5$ (left), $1.5$ (center), and $2.5$ (right).\label{bistab_ts_b}]{\includegraphics[width=0.33\linewidth]{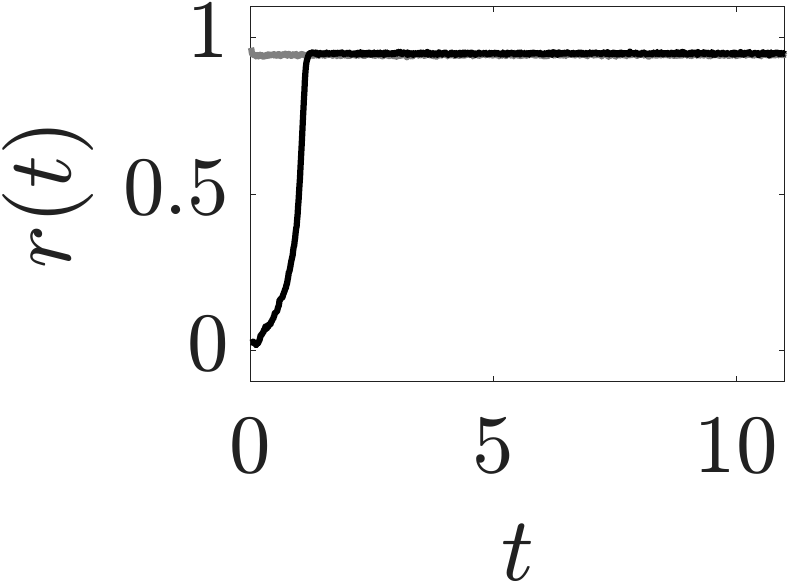}\includegraphics[width=0.33\linewidth]{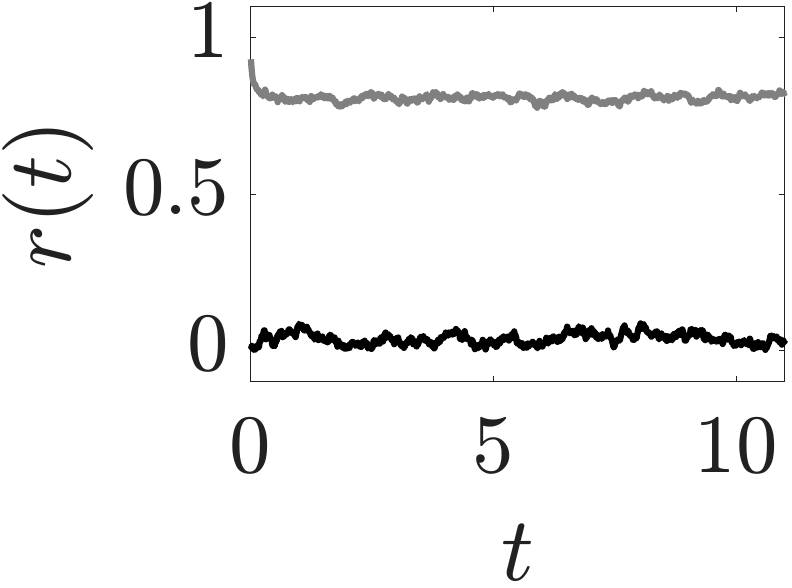}\includegraphics[width=0.33\linewidth]{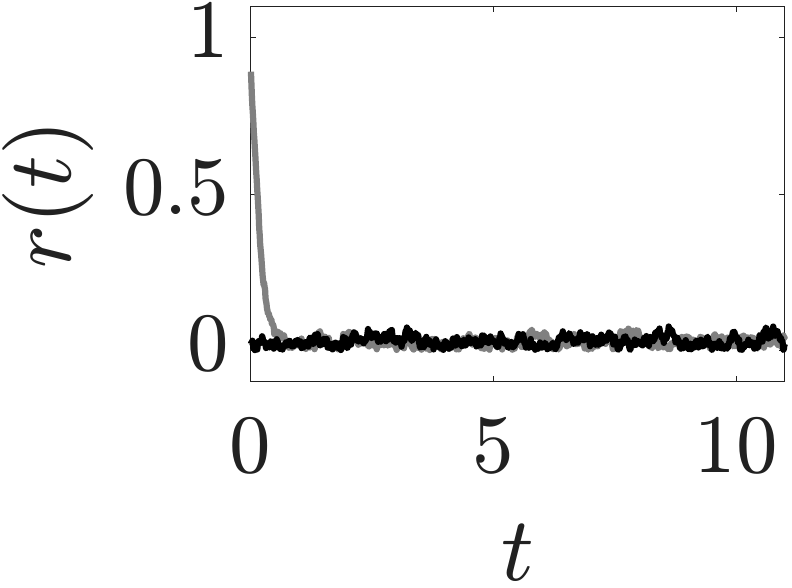}}
    \caption{(a)-(b): $r(t)$ as a function of $t$ for different parameter combinations by simulating Eqs.~\eqref{spatial-dynamics} and \eqref{phase-dynamics}. The blue (black) and red (gray) curves indicate the initial conditions $r(0)\approx 0.0$ and $\approx 1.0$, respectively. For all sub-figures, $(\gamma,\Omega,k)$ are identical to Fig.~\ref{phase-transition-common}.}
    \label{bistab_ts}
\end{figure}
To further illustrate key characteristics of bistable dynamics, Figs.~\ref{bistab_ts_a} and \ref{bistab_ts_b} show individual realisations of $r(t)$ across three qualitatively different dynamical regimes. Specifically, the trajectories corresponding to the monostable regimes are shown in the left (\ie, $\sigma^{\prime}_{\rm P}=0.1$ and $D^{\prime}_{\psi}=0.5$) and right panels (\ie, $\sigma^{\prime}_{\rm P}=1.9$ and $D^{\prime}_{\psi}=2.5$), where the dynamics converge to identical steady states irrespective of the initial conditions. This is in contrast to the bistable regime where the long-term dynamics depend strongly on the initial state and is demonstrated in the centre panels across both figures (\ie, $\sigma^{\prime}_{\rm P}=0.9$ and $D^{\prime}_{\psi}=1.5$). 
\begin{figure}[h]
    \centering
    \subfloat[$\Omega=1.0$; $\gamma=0.01$ (left), $0.1$ (center), and $10.0$ (right).\label{validate-GS-GM_a}]{\includegraphics[width=0.33\linewidth]{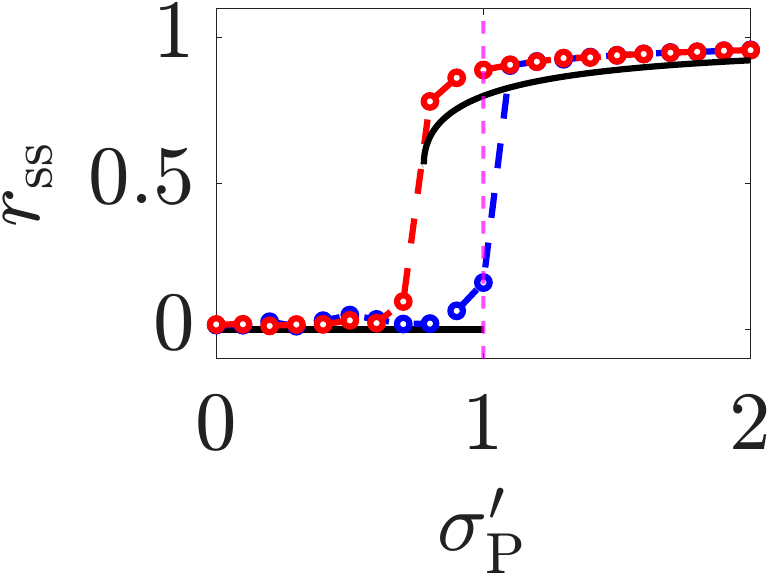}\includegraphics[width=0.33\linewidth]{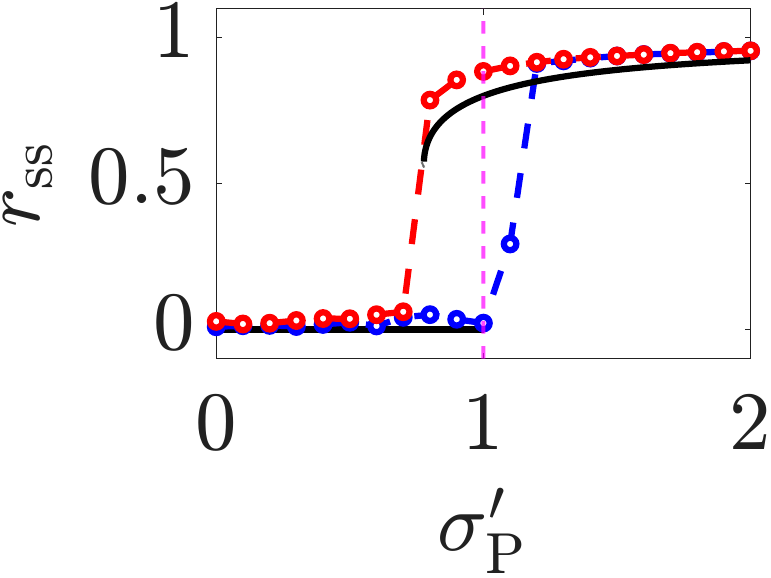}\includegraphics[width=0.33\linewidth]{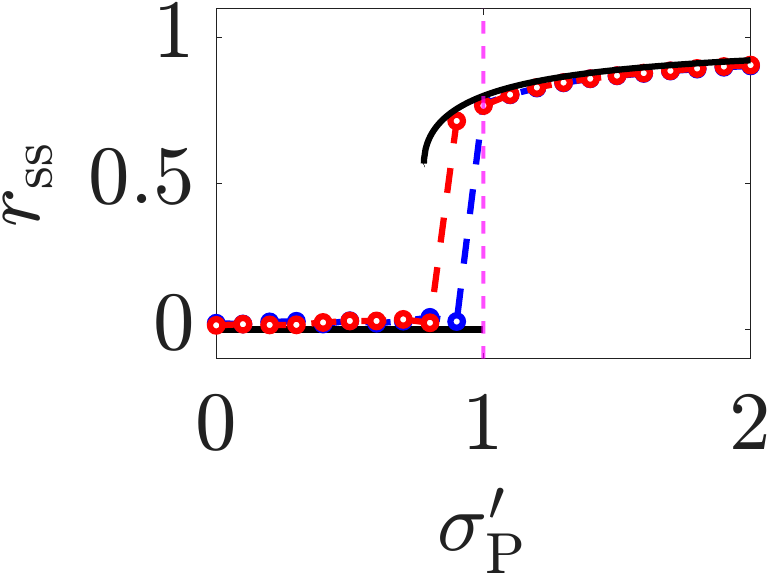}}

    \subfloat[$\gamma=1.0$; $\Omega=0.1$ (left), $1.0$ (center), and $10.0$ (right).\label{validate-GS-GM_b}]{\includegraphics[width=0.33\linewidth]{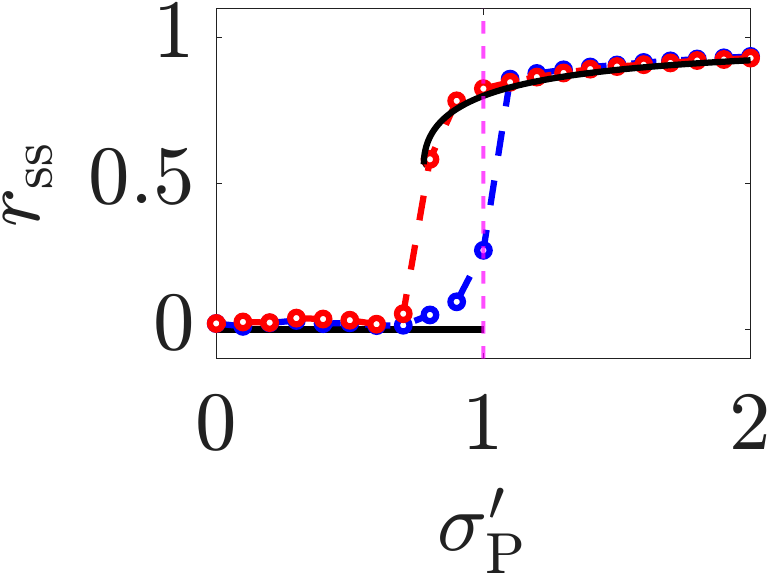}\includegraphics[width=0.33\linewidth]{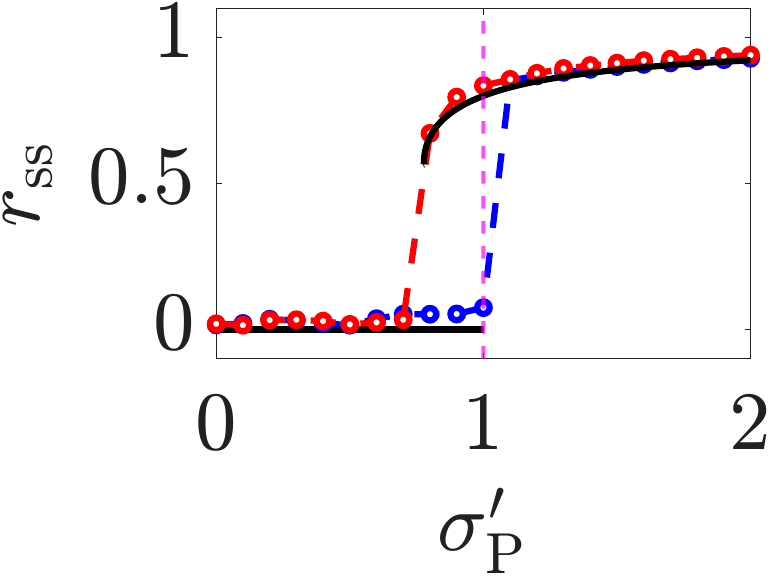}\includegraphics[width=0.33\linewidth]{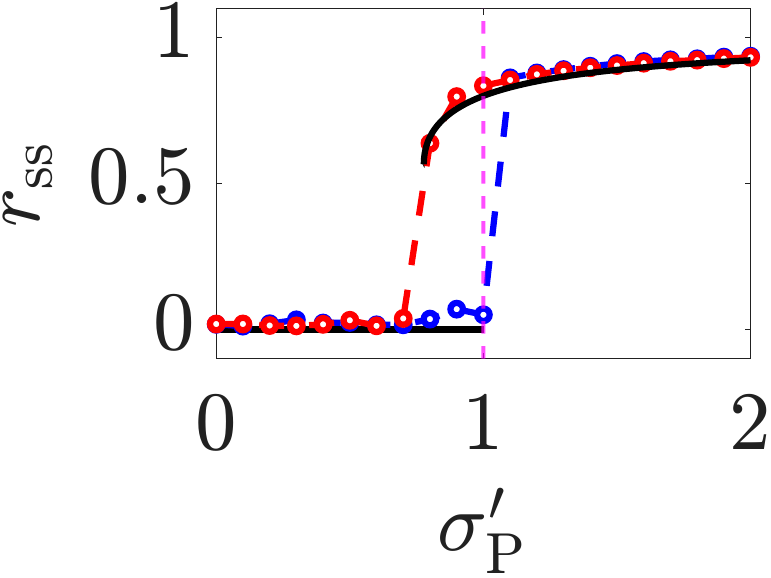}}

    \subfloat[$\Omega=\gamma=1.0$; $v_0=0.01$ (left), $0.10$ (center), and $10.00$ (right).\label{validate-GS-GM_c}]{\includegraphics[width=0.33\linewidth]{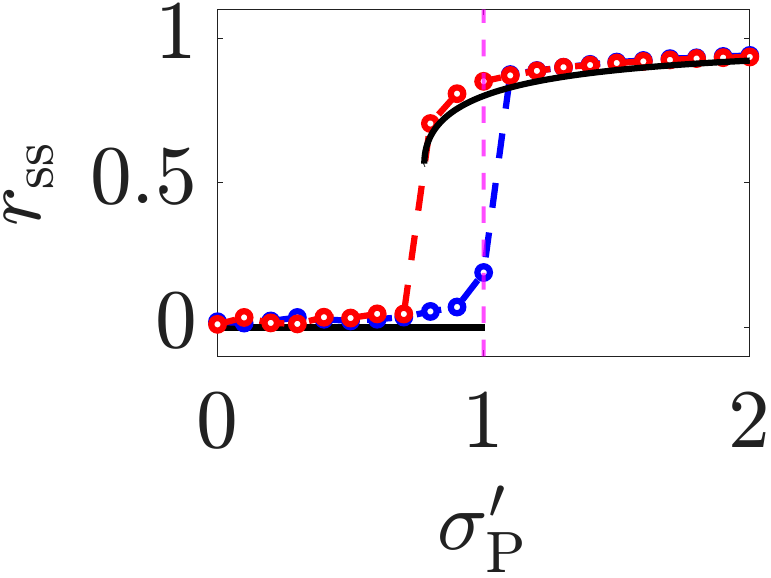}\includegraphics[width=0.33\linewidth]{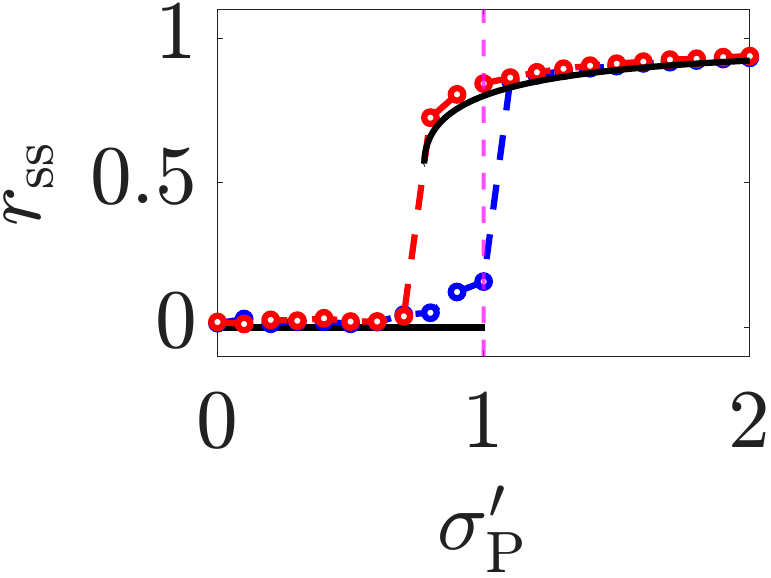}\includegraphics[width=0.33\linewidth]{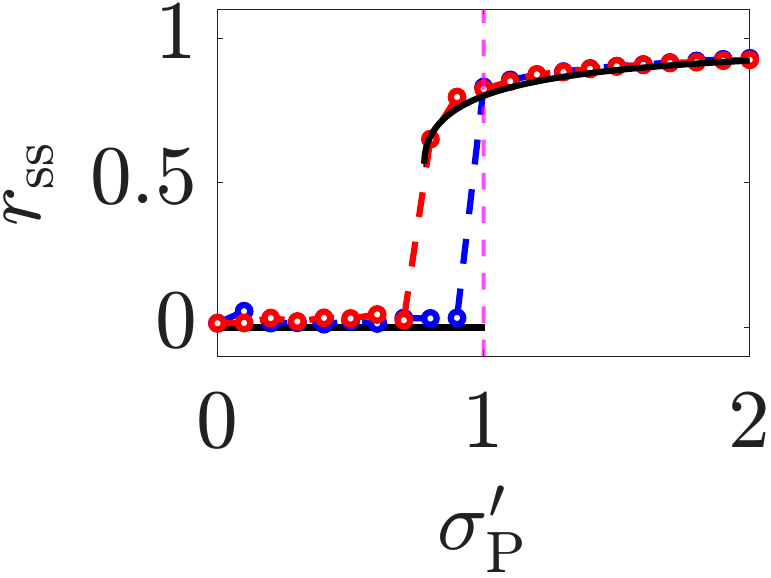}}

    \caption{(a)-(c): $r_{\rm ss}$ as a function of $\sigma^{\prime}_{\rm P}$ for different combinations of $\Omega,\gamma$, and $v_0$, respectively, whereas $(D,k,\sigma_{\rm R})=(1.0,2.0,5.0)$ for all.  Here, the values of $(T_{\rm R}, T_{\rm C})$, the meaning of the blue and red curves, as well as the vertical line, are identical to Fig.~\ref{phase-transition-common}.}
    \label{validate-GS-GM}
\end{figure}
Finally, Figs.~\ref{validate-GS-GM_a} and \ref{validate-GS-GM_b} demonstrate the phase transitions for multiple values of $(\Omega,\gamma)$ (\ie, parameters characterizing $G(\omega)$) and $v_0$, highlighting that the obtained predictions, especially that for $\sigma^{c}_{\rm P}$, remain reasonably accurate over a large parametric space across different orders of magnitude.

\begin{figure}[h]
    \centering

    \subfloat[$k=2.0$, $5.0$, $10.0$, and $15.0$ (left to right).\label{fig_k_vary_b}]{\includegraphics[width=0.245\linewidth]{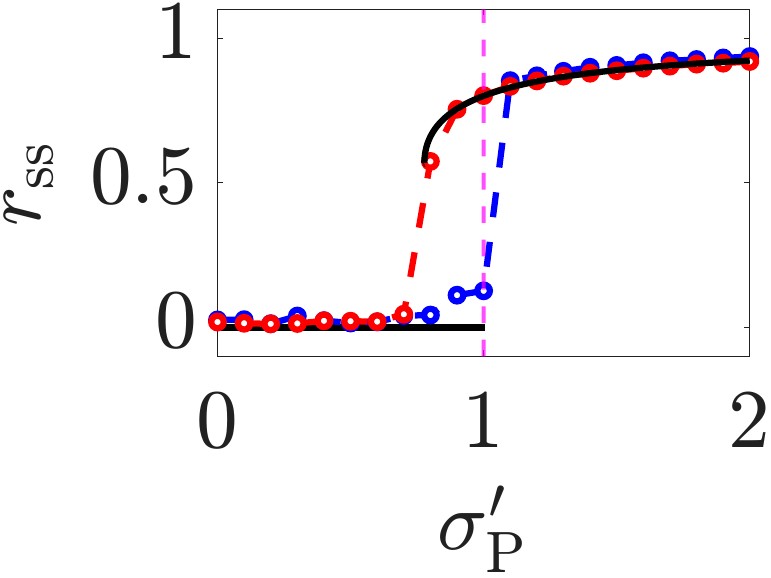}\includegraphics[width=0.245\linewidth]{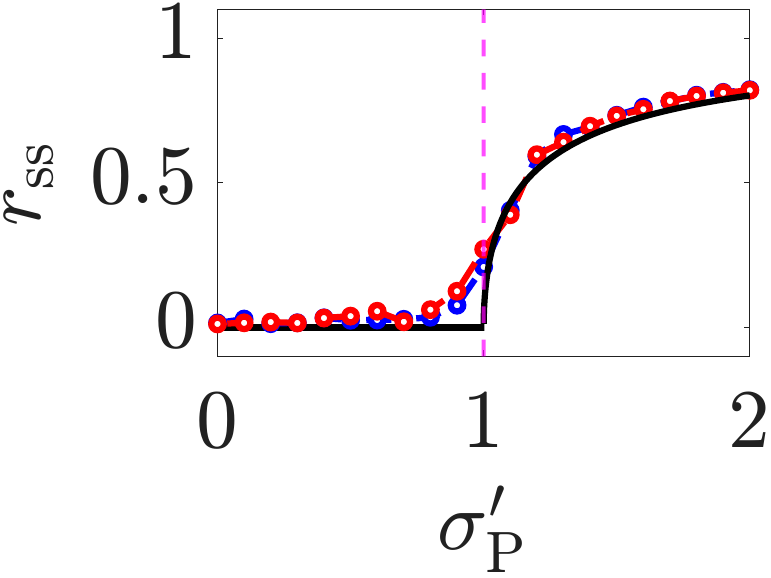}\includegraphics[width=0.245\linewidth]{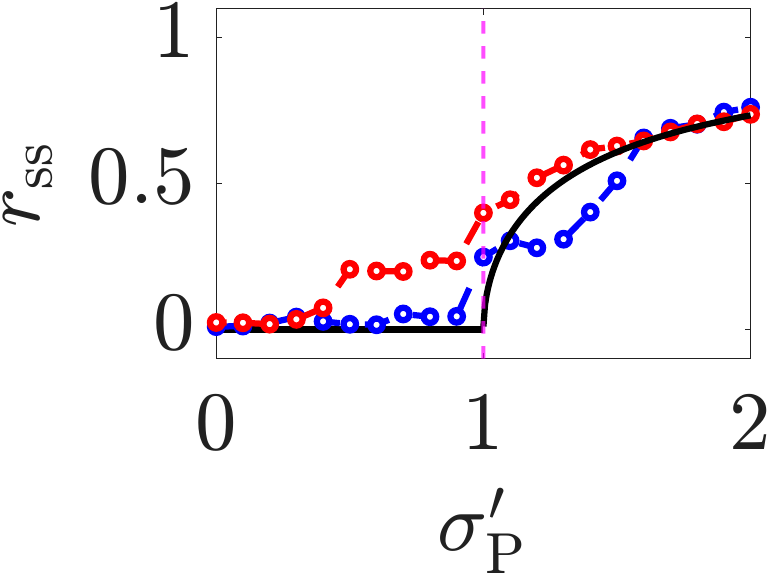}\includegraphics[width=0.245\linewidth]{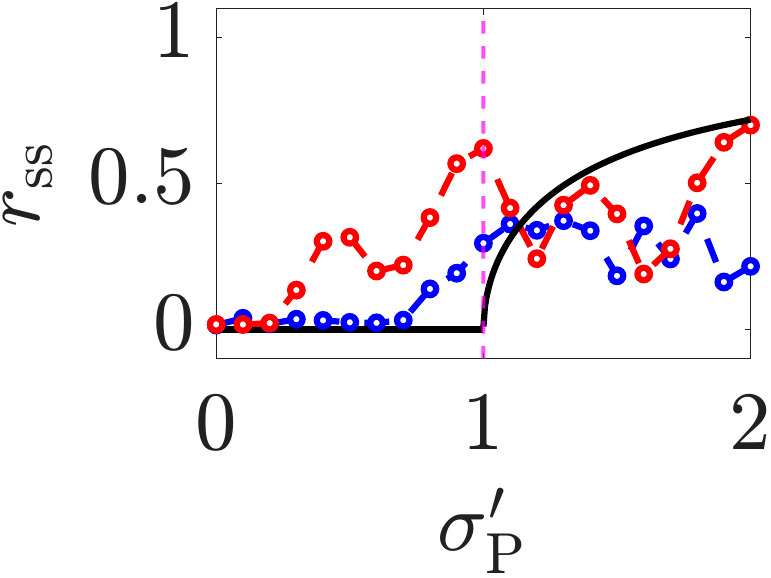}
    }

    \subfloat[$\sigma^{\prime}_{\rm P}=1.1$; $k=2.0$, $5.0$, $10.0$, and $15.0$ (left to right).\label{fig_k_vary_c}]{\includegraphics[width=0.248\linewidth]{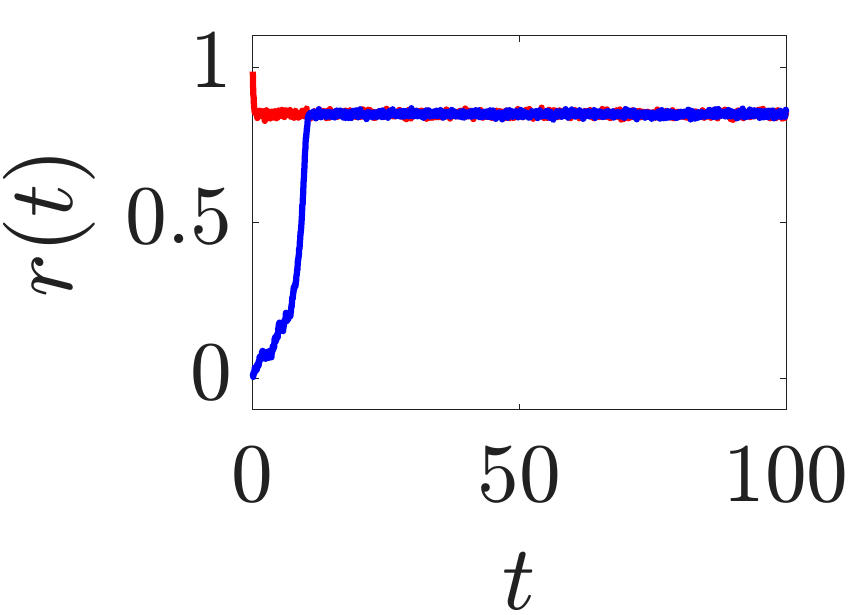}\includegraphics[width=0.248\linewidth]{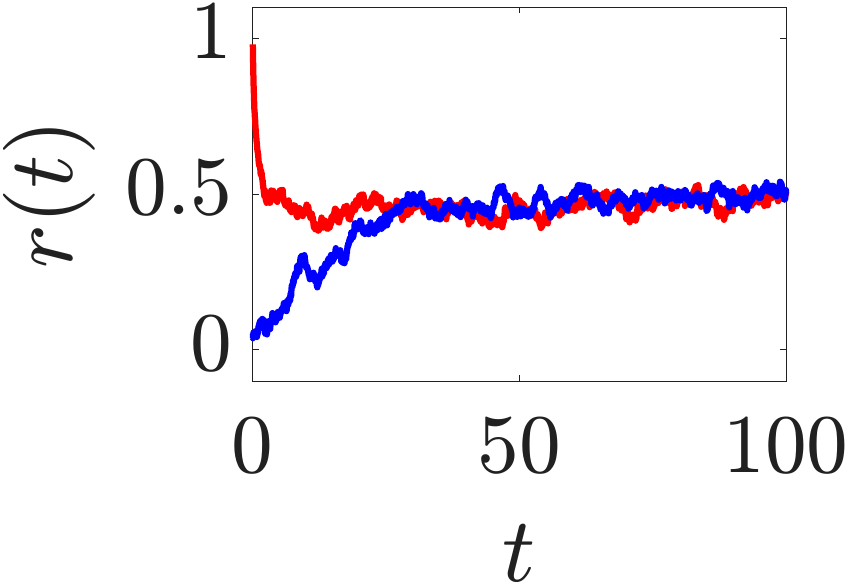}
    \includegraphics[width=0.248\linewidth]{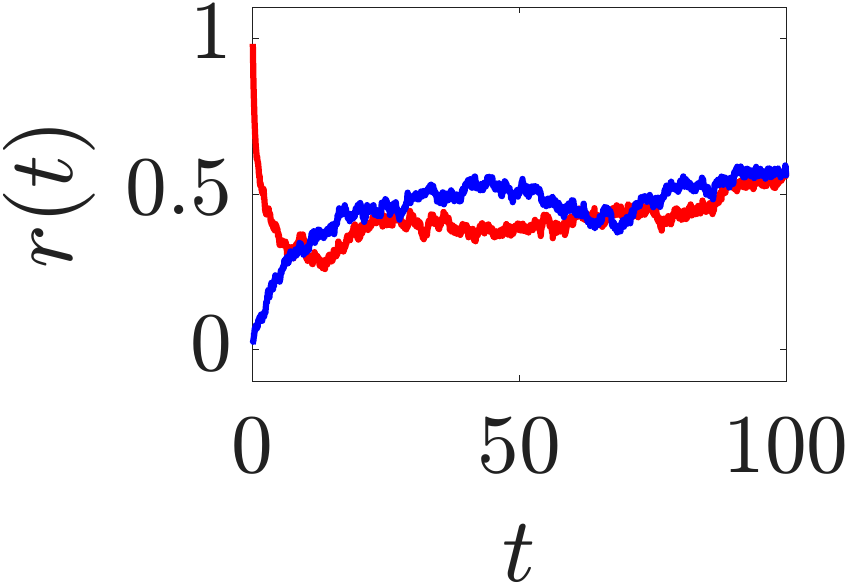}\includegraphics[width=0.248\linewidth]{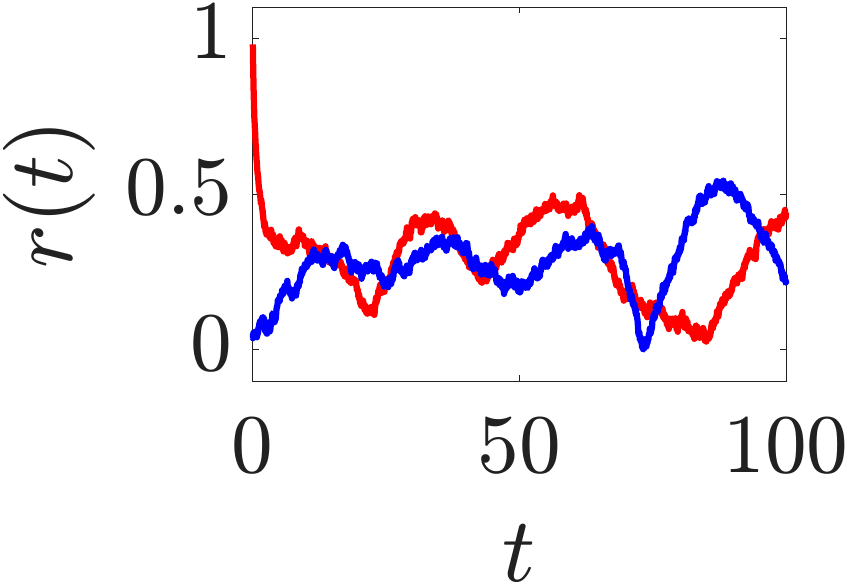}}

    \subfloat[$k=2.0$\label{fig_k_vary_d}]{\includegraphics[width=0.23\linewidth]{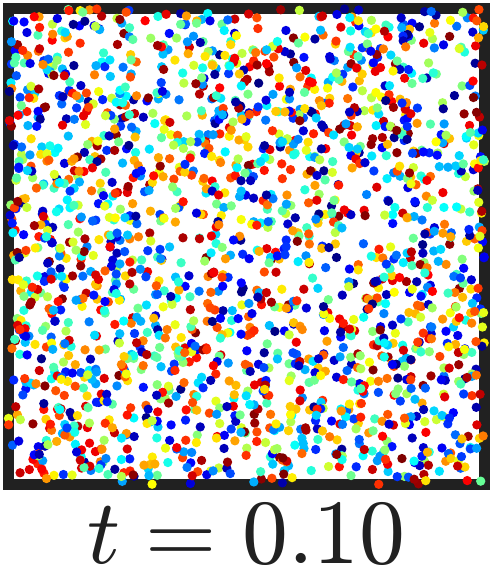}\hspace{1.5ex}\includegraphics[width=0.23\linewidth]{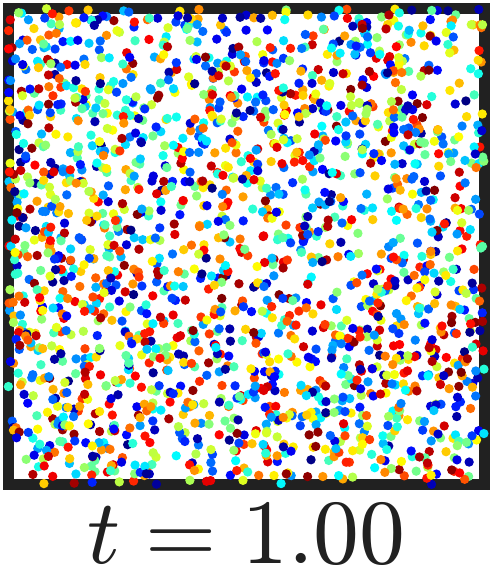}\hspace{1.5ex}\includegraphics[width=0.23\linewidth]{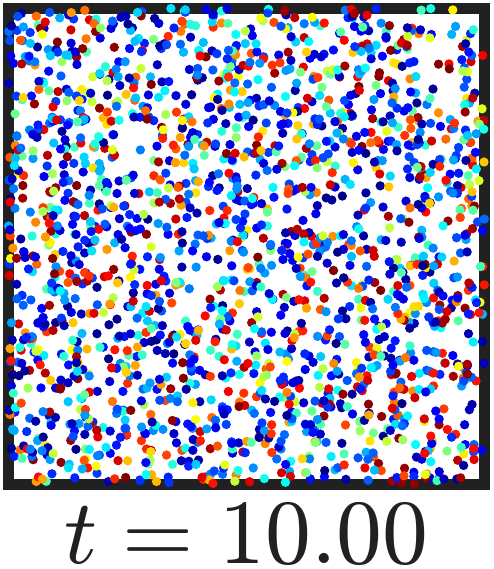}\hspace{1.5ex}\includegraphics[width=0.23\linewidth]{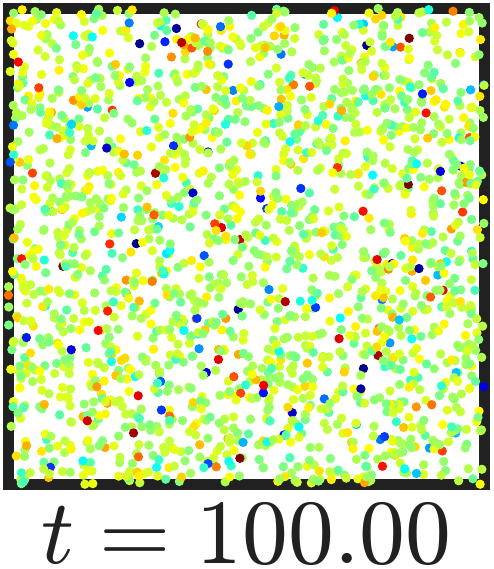}}

    \subfloat[$k=5.0$\label{fig_k_vary_e}]{\includegraphics[width=0.23\linewidth]{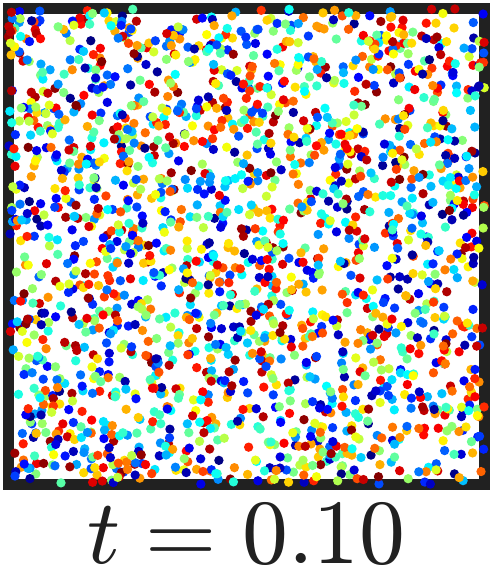}\hspace{1.5ex}\includegraphics[width=0.23\linewidth]{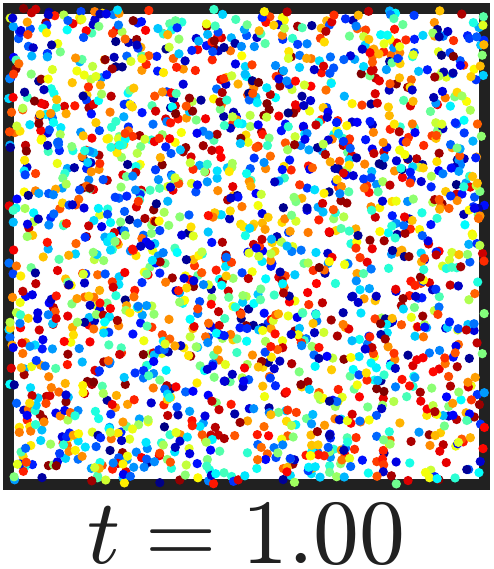}\hspace{1.5ex}\includegraphics[width=0.23\linewidth]{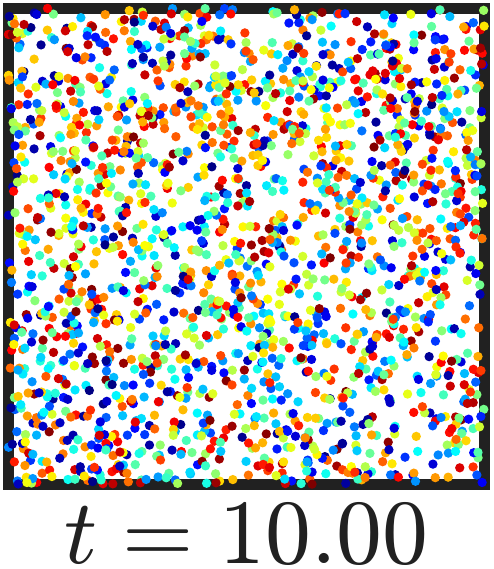}\hspace{1.5ex}\includegraphics[width=0.23\linewidth]{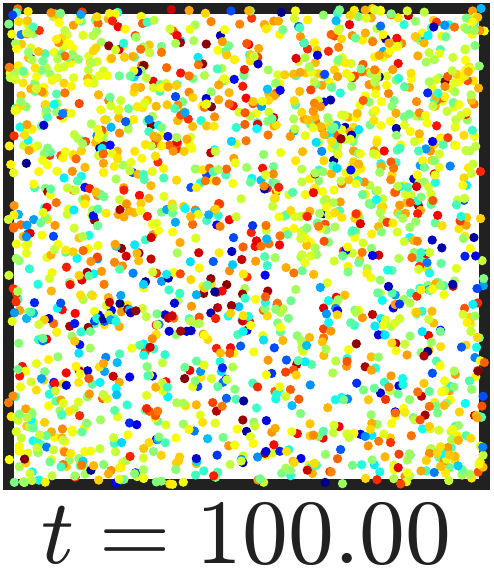}}

    \subfloat[$k=10.0$\label{fig_k_vary_f}]{\includegraphics[width=0.23\linewidth]{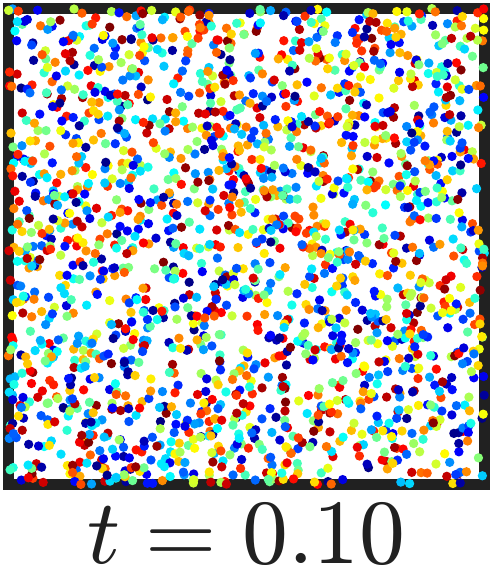}\hspace{1.5ex}\includegraphics[width=0.23\linewidth]{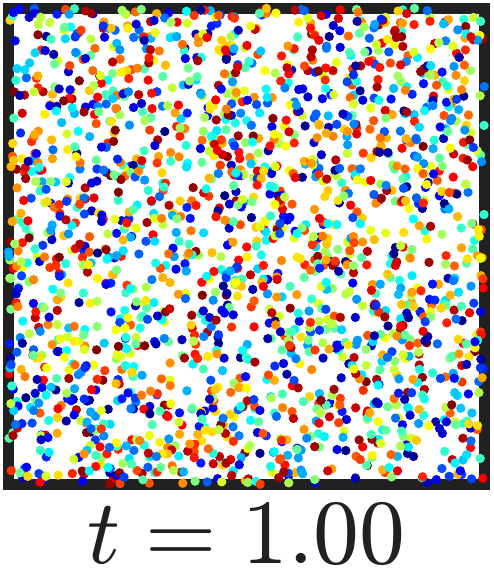}\hspace{1.5ex}\includegraphics[width=0.23\linewidth]{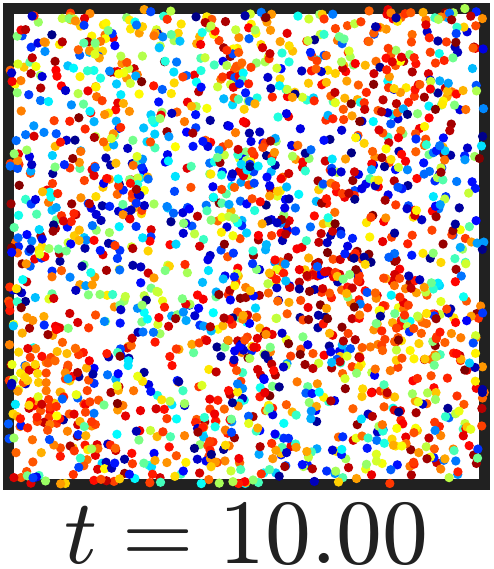}\hspace{1.5ex}\includegraphics[width=0.23\linewidth]{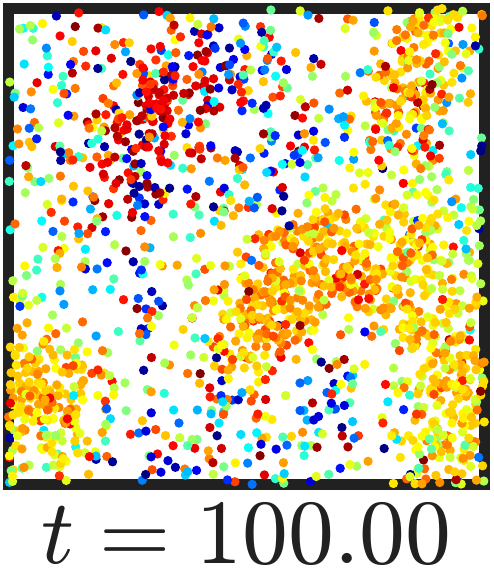}}

    \subfloat[$k=15.0$\label{fig_k_vary_g}]{\includegraphics[width=0.23\linewidth]{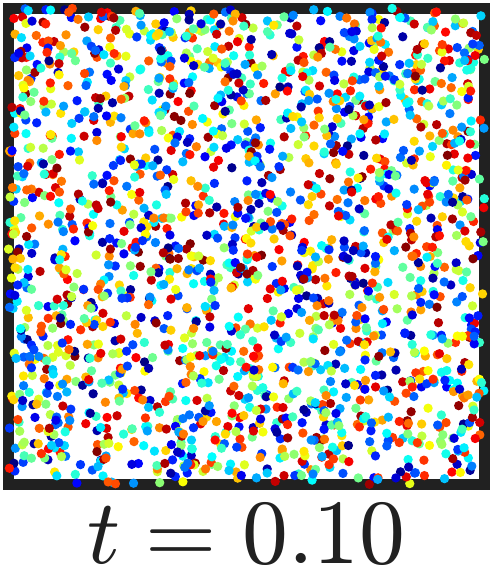}\hspace{1.5ex}\includegraphics[width=0.23\linewidth]{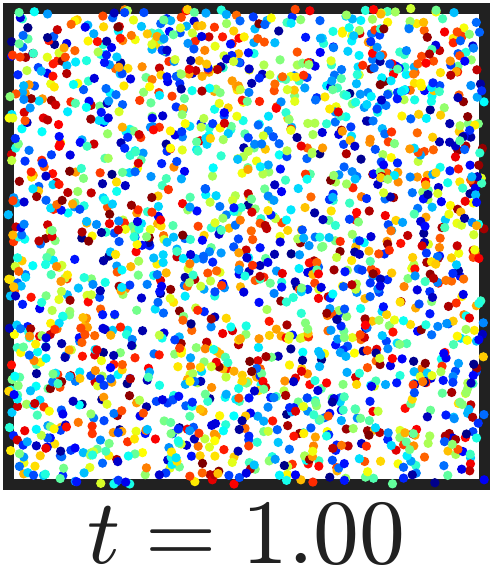}\hspace{1.5ex}\includegraphics[width=0.23\linewidth]{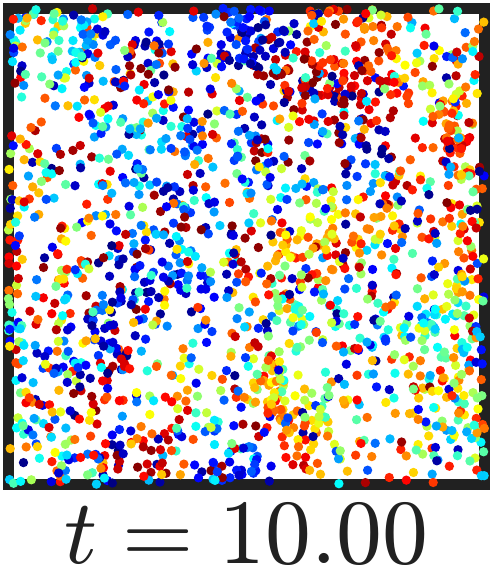}\hspace{1.5ex}\includegraphics[width=0.23\linewidth]{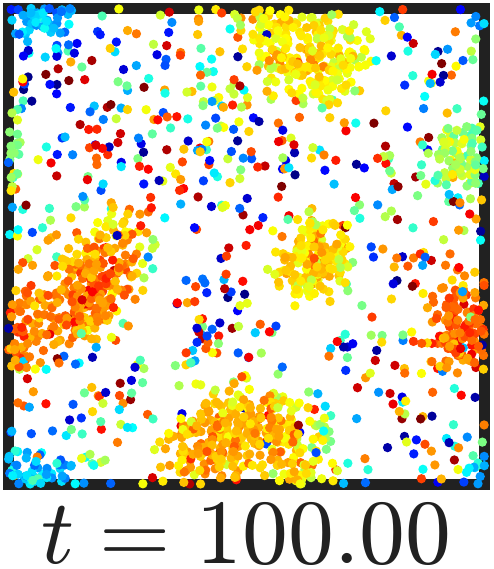}}
    
    \caption{(a): $r_{\rm ss}$ as a function of $\sigma^{\prime}_{\rm P}$ for different values of $k$.  Here, the values of $(T_{\rm R}, T_{\rm C})$, the meaning of the blue and red curves, as well as the vertical line, are identical to Fig.~\ref{phase-transition-common}; (b): $r(t)$ as a function of $t$ for $\sigma^{\prime}_{\rm P}=1.1$, obtained from Eqs.~\eqref{spatial-dynamics} and \eqref{phase-dynamics}, for different values of $k$; (c)-(f): Snapshots of the system, initialized asynchronously, for different values of $k$, whereas $\sigma^{\prime}_{\rm P}=1.1$ for all. Here, the markers represent the positions of the particles inside $\mathscr{L}$, whereas the marker colors indicate the instantaneous phases $\psi_i(t)\in [0,2\pi)$. Across all figures, the values of $(\Omega,\gamma,D,\sigma_{\rm R})=(1.0,1.0,0.5,5.0)$.}
    \label{fig_k_vary}
\end{figure}

While the observations across Figs.~\ref{phase-transition-common}-\ref{validate-GS-GM} are consistent with the predictions obtained in Sec.~\ref{math}, it is important to recall that the latter hinges critically on the assumption in Eq.~\eqref{ansatz}. However, the validity of this ansatz is not always guaranteed, warranting further investigation across broader parameter regimes to capture any deviations from the obtained theory. To this end, one of the key parameters is $k$ that controls the interaction range, the variation of which is shown in Fig.~\ref{fig_k_vary_b}. It is observed that increasing the value of $k$ (\ie, decreasing the interaction range) changes the nature of the phase transition from explosive to continuous. However, for very short-range interactions (here, $k\geq 10$), the obtained curves start to show substantial deviations from the predictions for both $\sigma^{c}_{\rm P}$ and $r_{\rm ss}$, in addition to the system not reaching a long-time steady state. The latter is demonstrated in Fig.~\ref{fig_k_vary_c}, which shows the unsteady and fluctuating order parameter dynamics for $\sigma^{\prime}_{\rm P}=1.1$, across both $k=10, 15$; this is in contrast to the steady order parameter dynamics achieved for both $k=2, 5$. These deviations can be attributed to the short interaction range, which leads to the formation of clustered states, violating the uniform spatial distribution assumption in Eq.~\eqref{ansatz}. This is visualized in Figs.~\ref{fig_k_vary_d}-\ref{fig_k_vary_g}, which show the formation of a spatially uniform state for $k=2,5$, whereas spatially clustered states arise for $k=10, 15$, with larger values of $k$ leading to tighter clusters. 
Taken together, the observations above highlight the critical role of the spatial distribution of particles in system dynamics and emergent phenomena such as synchronisation. 

\begin{figure}[h]
    \centering

    \subfloat[$c=4.0$, $8.0$, $16.0$, and $32.0$ (left to right).\label{init_vis}]{\includegraphics[width=0.245\linewidth]{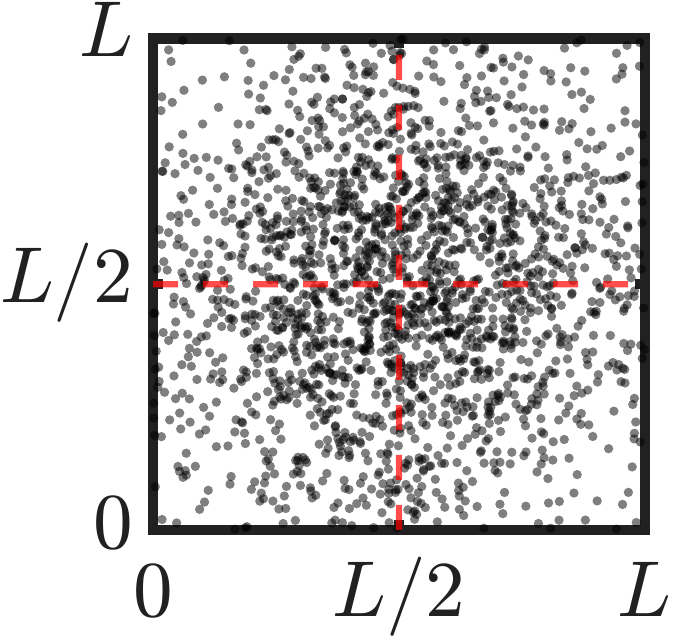}\includegraphics[width=0.245\linewidth]{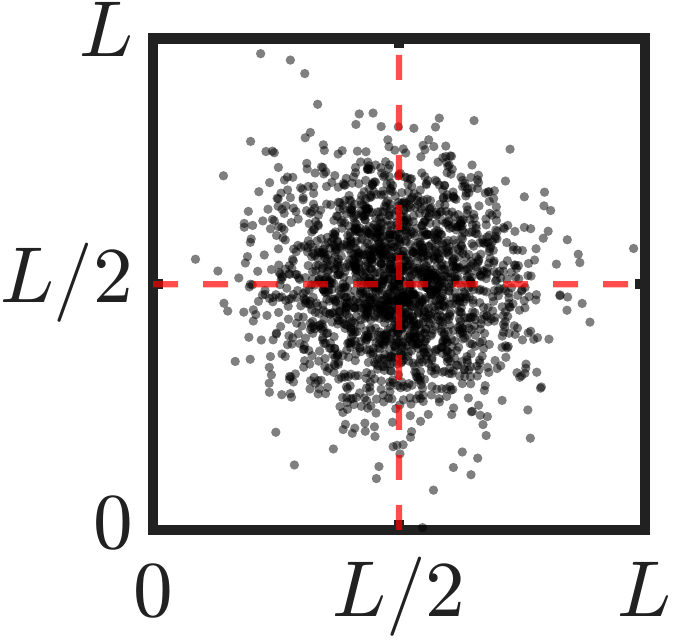}\includegraphics[width=0.245\linewidth]{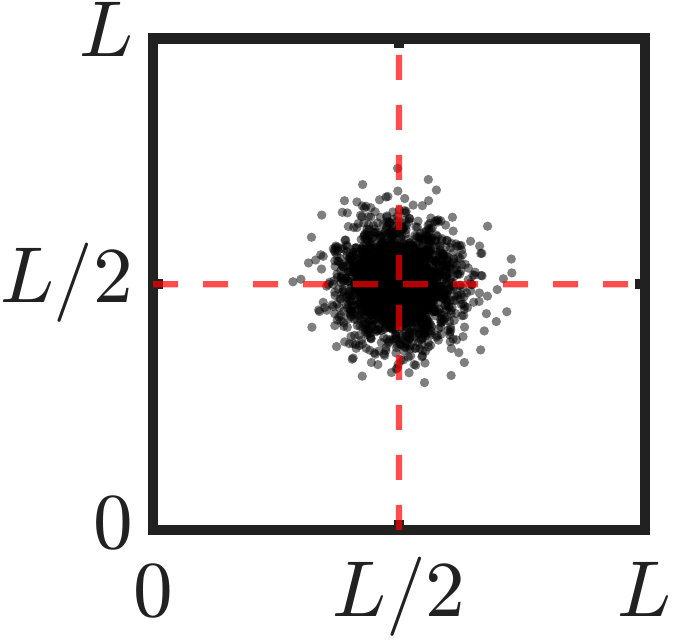}\includegraphics[width=0.245\linewidth]{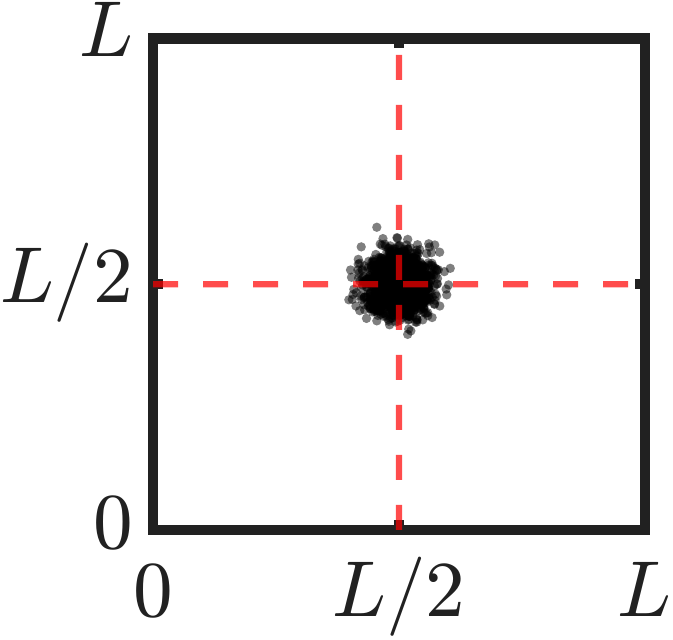}}
    
    \subfloat[$v_0=1.0$, $k=1$; $c=4.0$, $8.0$, $16.0$, and $32.0$ (left to right).\label{v0_1_c_vary}]{\includegraphics[width=0.245\linewidth]{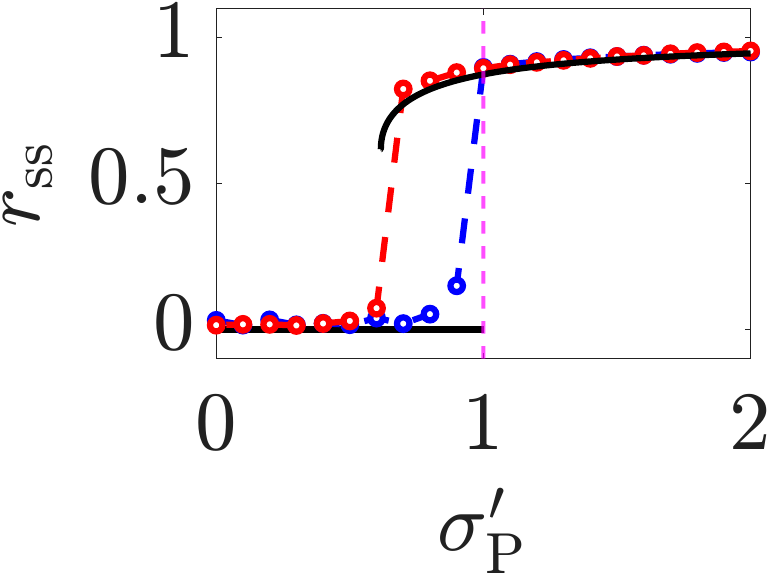}\includegraphics[width=0.245\linewidth]{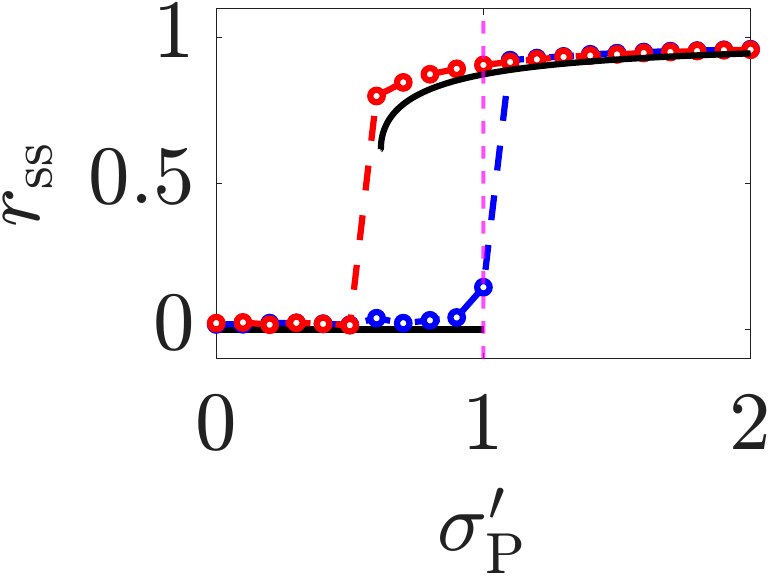}\includegraphics[width=0.245\linewidth]{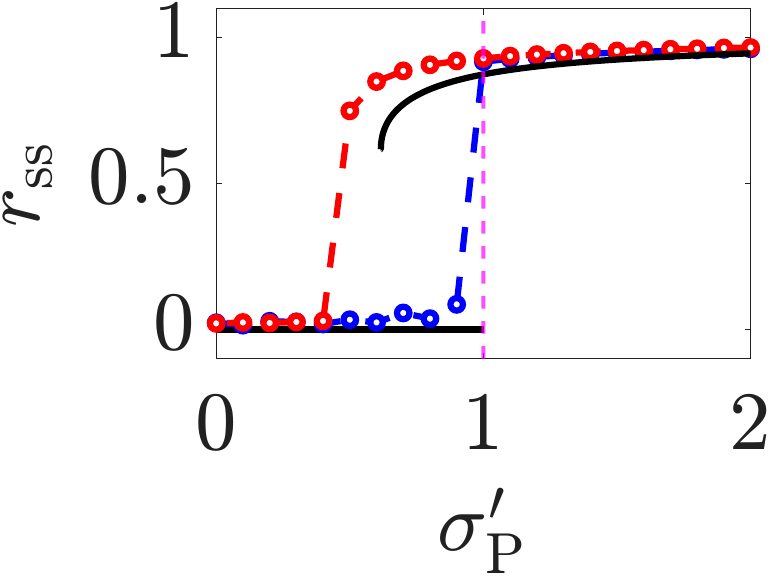}\includegraphics[width=0.245\linewidth]{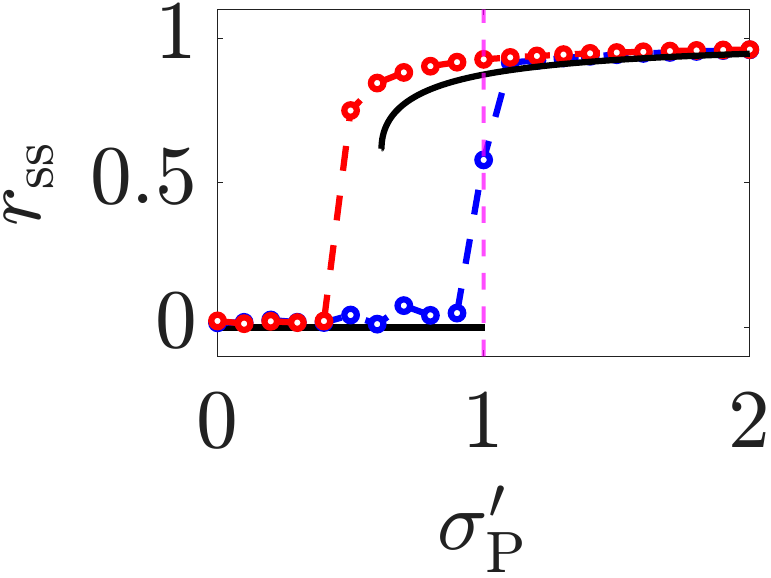}}

     \subfloat[$c=16.0$, $k=1$; $v_0=2.0$, $4.0$, $8.0$, and $16.0$ (left to right).\label{c_16_v0_vary}]{\includegraphics[width=0.245\linewidth]{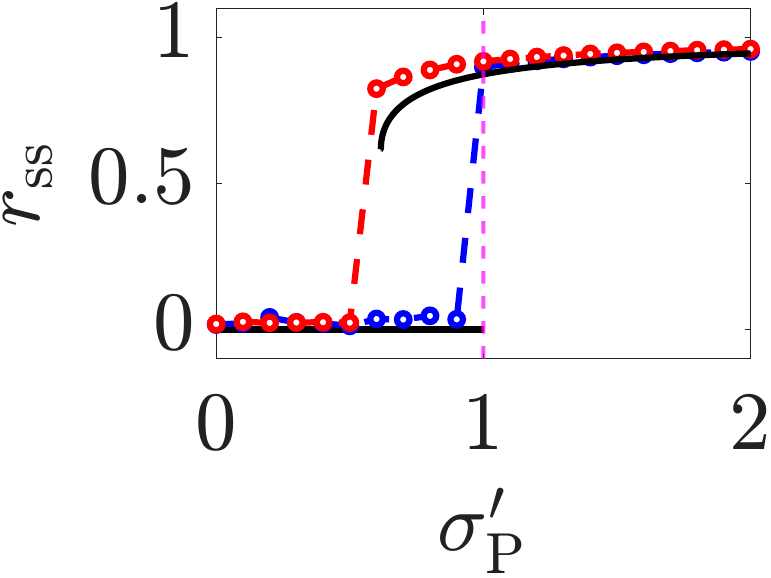}\includegraphics[width=0.245\linewidth]{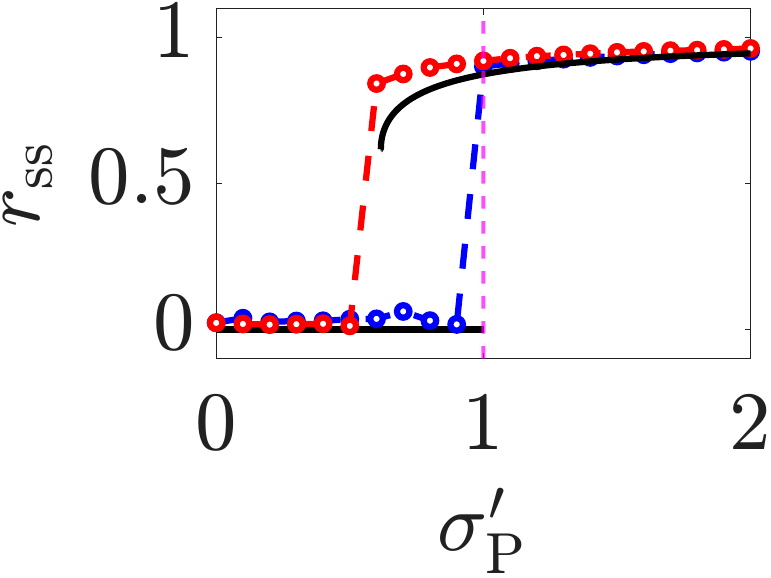}\includegraphics[width=0.245\linewidth]{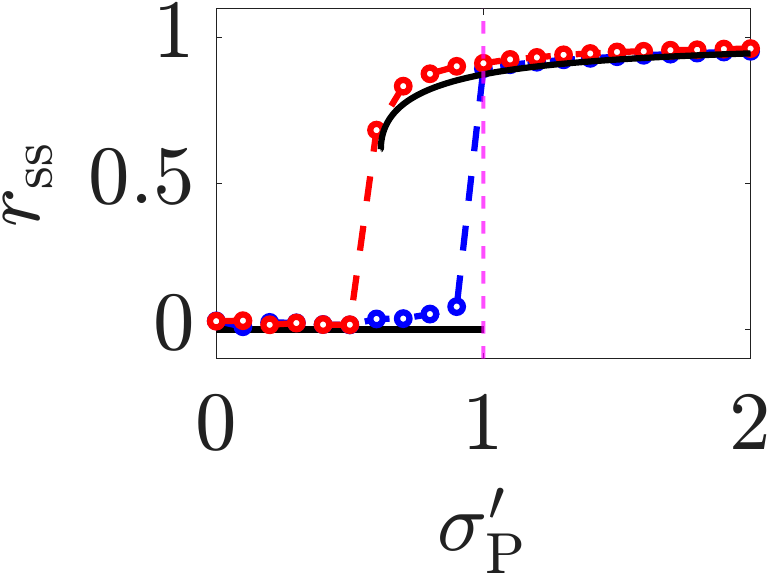}\includegraphics[width=0.245\linewidth]{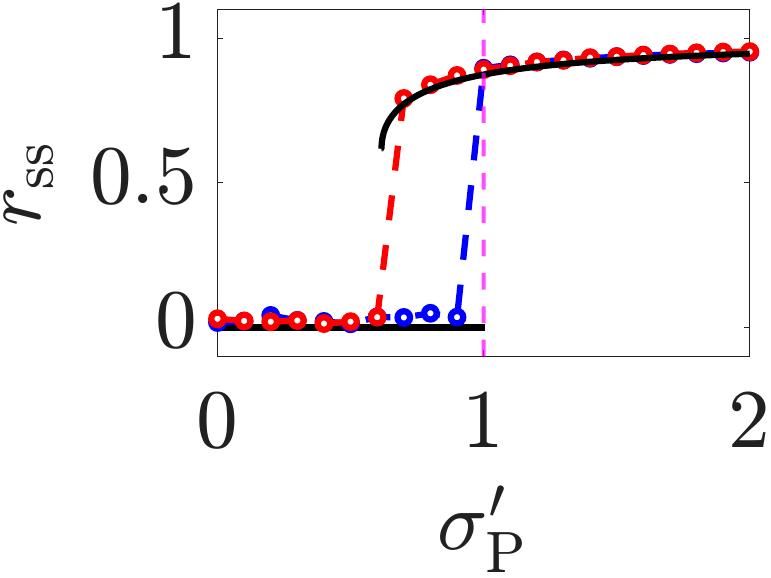}}

    \caption{(a): Initial spatial configuration of particles (using Eq.~\eqref{initial_spread}), for different values of $c$, where the particle positions are denoted by black markers; (b)-(c): $r_{\rm ss}$ as a function of $\sigma^{\prime}_{\rm P}$ for different combinations of $c$ and $v_0$. Here, the values of $(T_{\rm R}, T_{\rm C})$, the meaning of the blue and red curves, as well as the vertical line, are identical to Fig.~\ref{phase-transition-common}, whereas $(\Omega,\gamma, D,\sigma_{\rm R})=(1.0,1.0,1.0,5.0)$ across all figures.}
    \label{effect_c_v_vary}
\end{figure}

\begin{figure}[t]
    \centering

    \subfloat[$(c,v_0)=(16.0,1.0)$, $r(0)\approx 1.0$.\label{v0_1_c16_r0_1}]{\includegraphics[width=0.235\linewidth]{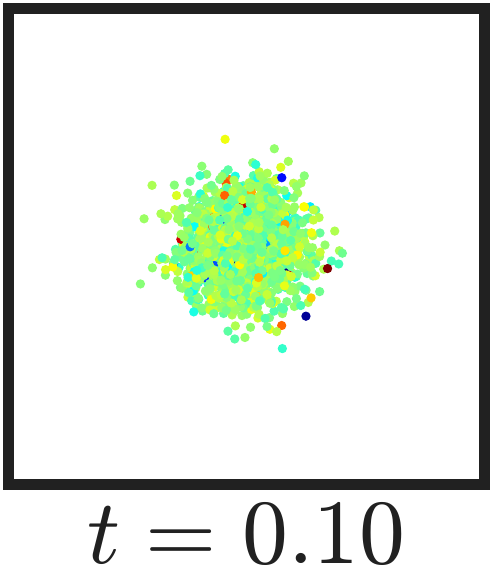}\hspace{1ex}\includegraphics[width=0.235\linewidth]{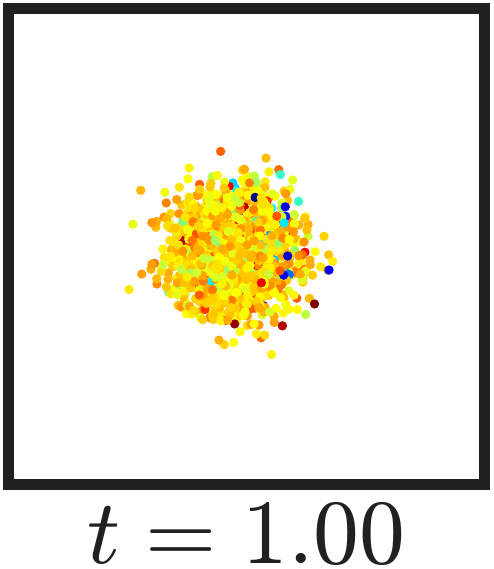}\hspace{1ex}\includegraphics[width=0.235\linewidth]{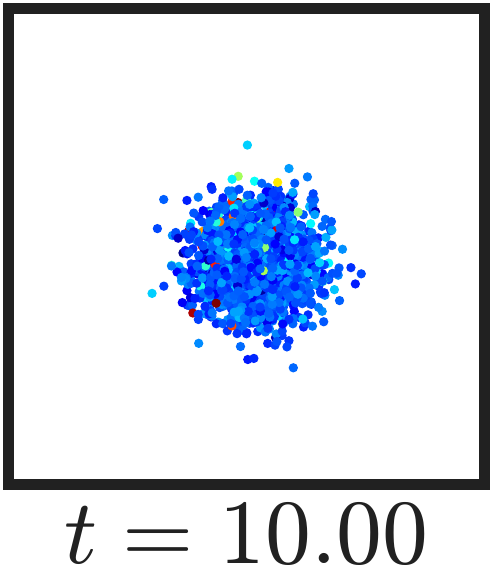}\hspace{1ex}\includegraphics[width=0.235\linewidth]{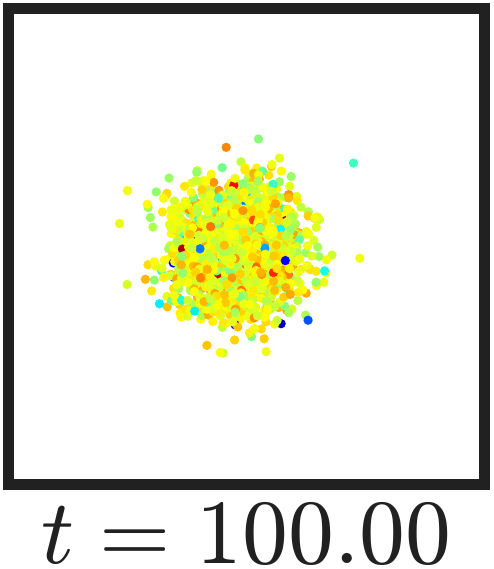}}

    \subfloat[$(c,v_0)=(16.0,1.0)$, $r(0)\approx 0.0$.\label{v0_1_c16_r0_0}]{\includegraphics[width=0.235\linewidth]{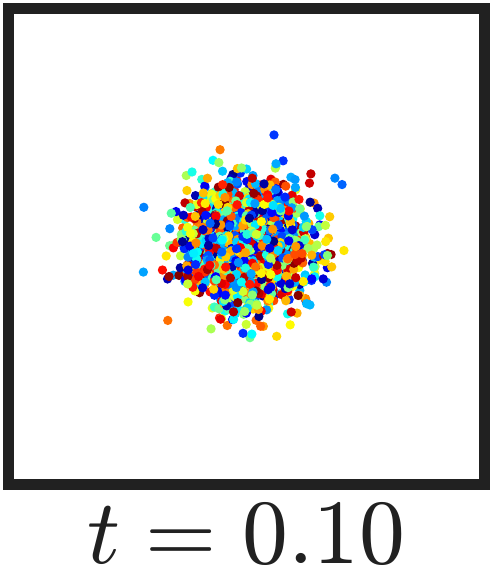}\hspace{1ex}\includegraphics[width=0.235\linewidth]{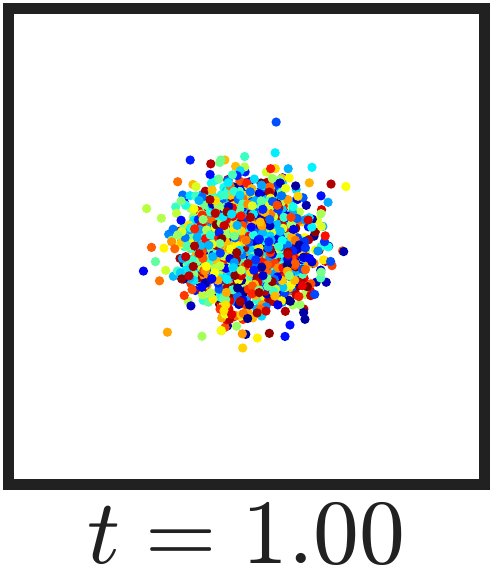}\hspace{1ex}\includegraphics[width=0.235\linewidth]{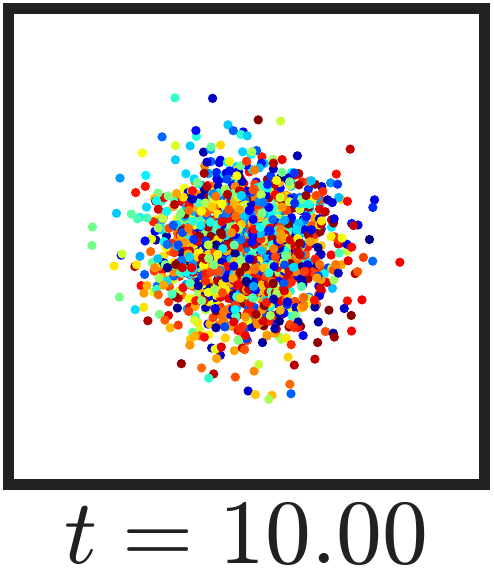}\hspace{1ex}\includegraphics[width=0.235\linewidth]{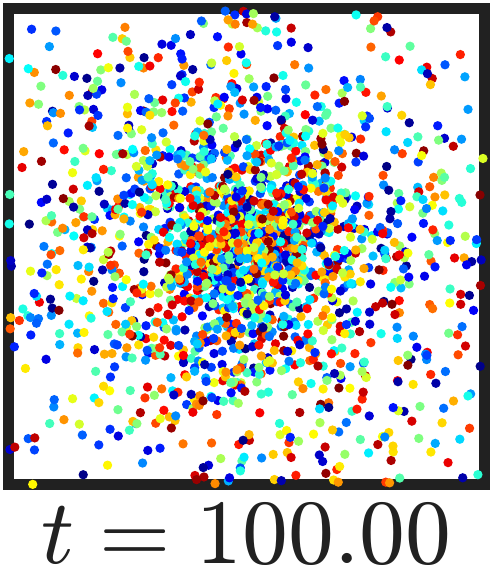}}

    \subfloat[$(c,v_0)=(16.0,16.0)$, $r(0)\approx 1.0$.\label{v0_16_c16_r0_1}]{\includegraphics[width=0.235\linewidth]{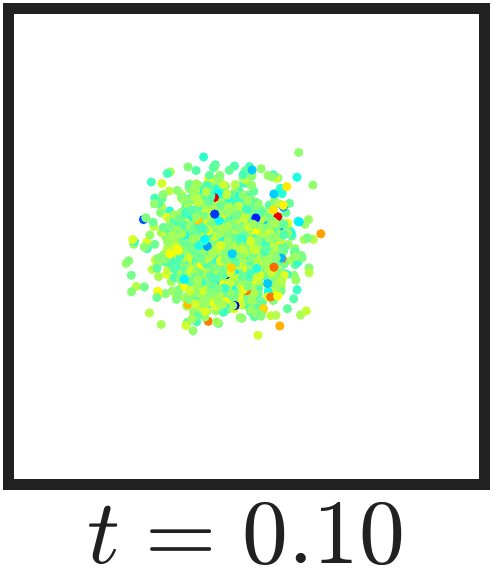}\hspace{1ex}\includegraphics[width=0.235\linewidth]{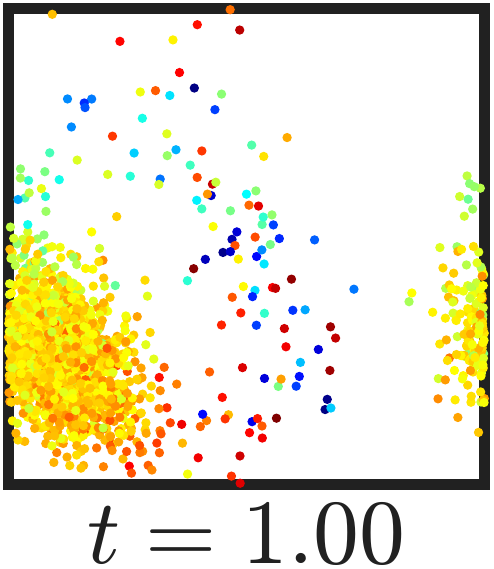}\hspace{1ex}\includegraphics[width=0.235\linewidth]{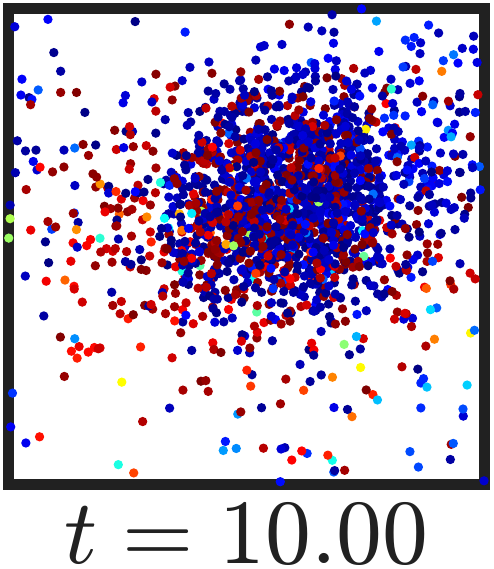}\hspace{1ex}\includegraphics[width=0.235\linewidth]{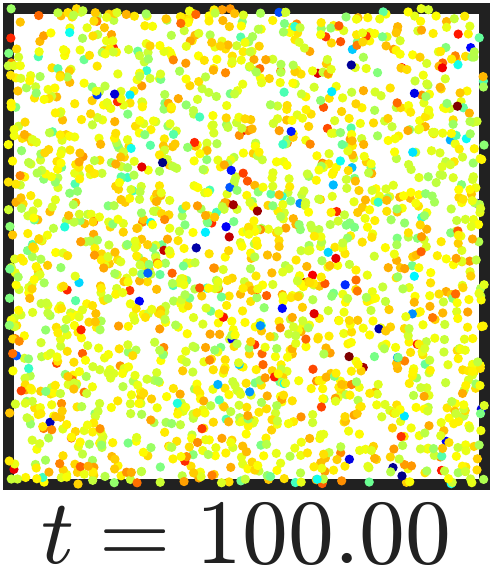}}

    \subfloat[$(c,v_0)=(16.0,16.0)$, $r(0)\approx 0.0$.\label{v0_16_c16_r0_0}]{\includegraphics[width=0.235\linewidth]{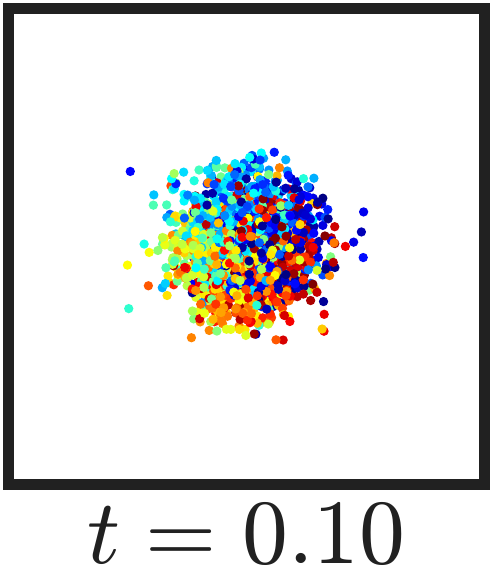}\hspace{1ex}\includegraphics[width=0.235\linewidth]{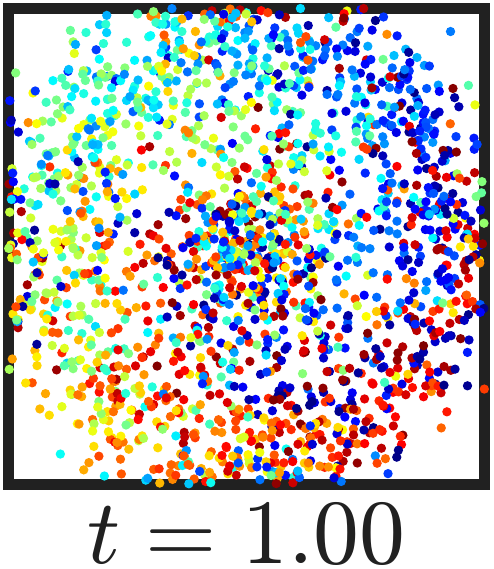}\hspace{1ex}\includegraphics[width=0.235\linewidth]{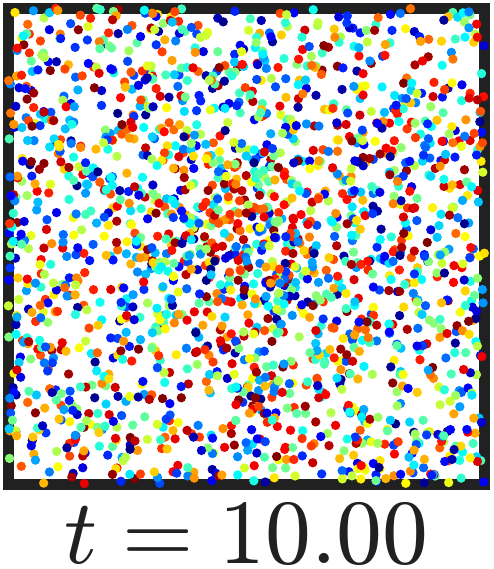}\hspace{1ex}\includegraphics[width=0.235\linewidth]{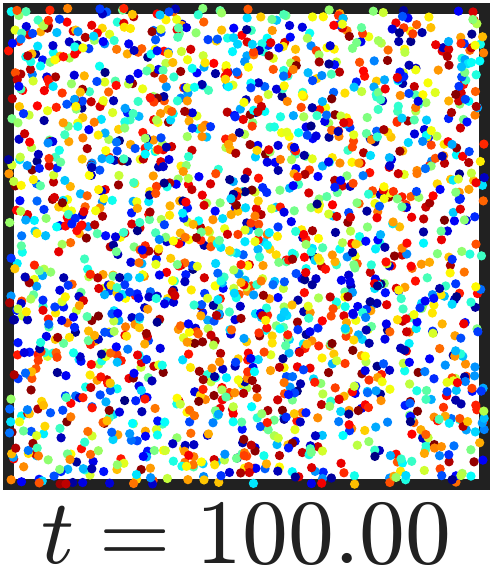}}
    
    \caption{(a)-(d): Snapshots of the system configuration for different combinations of $c$, $v_0$, and $r(0)$, at different time instants, whereas $\sigma^{\prime}_{\rm P}=0.80$ for all. The colors and points have the same meaning as in Fig.~\ref{fig_k_vary_d}, whereas the values of the constants $(k,\Omega,\gamma,D,\sigma_{\rm R})$ are identical to that of Fig.~\ref{v0_1_c_vary}.}
    \label{fig_cv_vary_snaps}
\end{figure}

The above observations also suggest that the initial spatial distribution of the particles is expected to play a key role in determining the system dynamics. To investigate this, the particle positions at $t=0$ are assigned following
\begin{equation}
    x_i(0)=(L/2)+U_i,\quad
    y_i(0)=(L/2)+V_i.
\label{initial_spread}
\end{equation}
$\forall\ i=1,2,\dots,N$. Here, $(x_i(0), y_i(0))$ are the initial particle coordinates which are computed such that they lie within $\mathscr{L}$, whereas $(U_i, V_i)$ are Gaussian-distributed with zero mean and standard deviation $\lambda$. The latter is defined in terms of the domain size $L$ as $\lambda=L/c$, such that larger values of $c$ imply more tightly clustered configurations centred at $(L/2,L/2)$; see Fig.~\ref{init_vis} for visualisations of the initial system configuration at different values of $c$.
Subsequently, Fig.~\ref{v0_1_c_vary} demonstrate the phase transitions for different values of $c$. Curiously, it is observed that, while the numerically obtained phase transition point remains consistent with the theoretical prediction, the value of the steady-state on the synchronous branch deviates from the prediction, especially in the bistable regime, with increasing values of $c$; in particular, the numerically obtained value of $r_{\rm ss}$ is seen to be slightly larger than its theoretically predicted value. However, upon increasing the propulsion velocity $v_0$, it is observed that this discrepancy disappears, with near-perfect agreement observed for $v_0=16.0$; see Fig.~\ref{c_16_v0_vary}. Qualitatively, this behavior can be attributed to the fact that at lower velocities, the synchronized initial state supports a higher magnitude of steady-state synchronization, whereas the asynchronous state spreads out similar to non-clustered distributed state (see Figs.~\ref{v0_1_c16_r0_1}-\ref{v0_1_c16_r0_0}); however, at higher velocities, both initial configurations spread out rapidly to fill the available domain (see Figs.~\ref{v0_16_c16_r0_1}-\ref{v0_16_c16_r0_0}), which ensures that the spatial distribution follows the ansatz in Eq.~\eqref{ansatz}, leading to much better agreement between theory and numerics compared to the previous case. 

\section{Summary and Outlook}
\label{conc}

To summarize, this paper explores the effect of distance-dependent higher-order interactions in chiral active particles, where the dynamics are modeled through an analytically tractable Vicsek-Kuramoto system that incorporates coupled spatial and alignment dynamics. The higher-order interactions are introduced via triadic (\ie, three-body) couplings, with their spatial dependence captured by a function that more closely approximates realistic interparticle interactions. For the purposes of analysis, the original model is mapped onto an effective Kuramoto framework with effective coupling strengths using a spatial uniformity ansatz, yielding analytical predictions for critical phase-transition points and approximate steady-state order-parameter values. The obtained expressions reveal that while the critical pairwise coupling threshold increases with frequency heterogeneity $\gamma$, angular noise $D_{\psi}$, and the interaction range parameter $k$, it remains independent of the triadic coupling strength $\sigma_{\rm T}$, with similar observations for the critical noise threshold $D^c_{\psi}$ as well. Qualitatively, triadic couplings fundamentally alter the nature of the phase transition, transforming a continuous transition observed in the purely pairwise case into an explosive transition, characterized by the formation of a bistable parameter regime near the critical point and a hysteresis loop whose width expands with increasing triadic interaction strength. 

Numerical simulations of the governing equations show good agreement with theoretical predictions across broad parameter regimes, accurately capturing both classes of phase transitions and the post-transition steady-state dynamics. The simulations also highlight the limits of the spatial uniformity assumption, where the introduction of 
very short-range interactions leads to the formation of spatial clusters that violate the spatial homogeneity ansatz. This causes the phase transition to revert from explosive to continuous and induces non-stationary fluctuations in the order parameter dynamics. Furthermore, localized initial spatial configurations produce deviations in the steady-state synchronisation level along the synchronous branch; however, increasing the self-propulsion velocity rapidly disperses clusters across the domain, thereby re-establishing spatial homogeneity and restoring quantitative agreement between theory and simulations. Collectively, these findings provide a foundational theoretical framework for understanding higher-order alignment mechanisms in active matter systems, with broad relevance to micro-swimmers, biological swarms, and synthetic agents.

Building on the results of this study, several promising directions for future research on active matter systems also emerge. These include, but are not limited to, extending the spatial interaction to include spherical geometries~\cite{PhysRevE.91.022306} and channel confinements~\cite{PhysRevE.108.034605} to reveal how surface curvature and boundary constraints interplay with higher-order alignment mechanisms, introducing adaptive interactions~\cite{chen2024synchronization,PhysRevE.109.024221} where coupling strengths evolve dynamically with local phase coherence and/or mechanical feedback to uncover self-organization principles, and exploring combined attractive-repulsive interactions~\cite{frolov2020chimera} and extending multi-body couplings beyond triadic terms to yield novel multi-stable regimes and classes of dynamical transitions.

\begin{acknowledgments}
The authors acknowledge IIT Kharagpur for providing computational and other research facilities. DB acknowledges financial support from IIT Kharagpur via an Institute Postdoctoral Fellowship (No: IIT/PDF/2025/PH/14). PSB acknowledges financial support from ANRF, Government of India, project-wide CRG/2023/006186. PP acknowledges financial support from IIT Kharagpur via the Faculty Research Scheme for Research Grant (No: IIT/SRIC/PH/RUS/2024-2025/065).
\end{acknowledgments}

\section*{Data Availability Statement}

No empirical data were created during the study. The computer programs and scripts developed during the study are available on request from the corresponding author.

\section*{Conflict of Interest}
The authors have no conflicts to disclose.

\appendix

\section{Evaluating the integral $I_i$}
\label{critical-integral}

The integral involved in both Eqs.~\eqref{Z1_i} and \eqref{Z2_i} is
\begin{equation}
I_i=\int_{\mathscr{L}} \D^2\bm{r}_{j_1}\ F\left(d_{ij_1}\right),\label{difficult-integral}
\end{equation}
where $\mathscr{L}$ denotes a square box of side $L$ with periodic boundary conditions (PBCs) on each side, and $d_{ij_1}$ is the distance between the reference particle situated at $x_i$ and other particles at $x_j$. This can be expressed as
\begin{equation}
I_i=\int_{0}^L \D x_{j_1}\int_{0}^L \D y_{j_1} \exp\left[-\frac{\left(x_{j_1}-x_i\right)^2+\left(y_{j_1}-y_i\right)^2}{d^2_0}\right]=\mathscr{A}^2_i,
\end{equation}
since the integrand can be factored into two identical components.
Thus, to evaluate the integral in Eq.~\eqref{difficult-integral}, it is sufficient to calculate the integral
\begin{equation}
    \mathscr{A}_i=\int_{0}^{L} \D x_j\ \exp\left[-\left(x_j-x_i\right)^2/d^2_0\right],
\label{1d-integral}
\end{equation}
where the PBCs on the domain $[0, L)$ of the system require that distances be computed using the minimum image convention (MIC) such that $(x_j-x_i)^2\leq L/2$. 
\begin{figure}[h]
    \centering
    \includegraphics[width=0.8\linewidth]{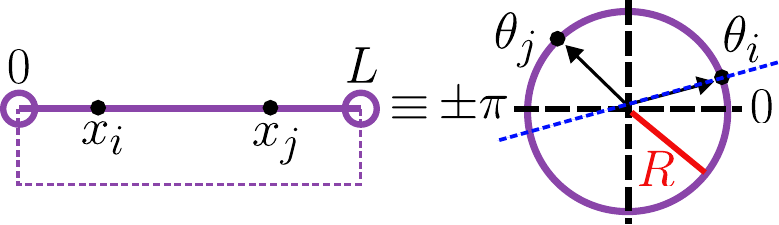}
    \caption{Schematic representation of (left) the domain between $x=0$ and $x=L$ over which the integral in Eq.~\eqref{1d-integral} is evaluated, where the broken line highlights its periodic nature; and (right) the equivalent circular topology of radius $R=L/2\pi$, where the positions are expressed in terms of $\theta\in[-\pi,\pi)$. Here, $(x_i,\theta_i)$ and $(x_j,\theta_j)$ are the reference position and dummy positions over which the integral is evaluated, respectively, with the angles being measured from the positive x-axis. The blue broken line indicates that the circular topology can be rotated such that $\theta_i=0$ without loss of generality.}
    \label{integral-image}
\end{figure}
Now, a line of length $L$ with PBCs is equivalent to a circle of radius $R=L/2\pi$, where the position of each particle $k$ can be parameterised by the angle $\theta_k$ such that $x_k=R\theta_k=L\theta_k/2\pi$; see Fig.~\ref{integral-image} for a schematic. In this setup, the integral in Eq.~\eqref{1d-integral} takes the form
\begin{equation}
    \mathscr{A}_i=R\int_{-\pi}^{+\pi} \D\theta_j\ \exp\left(-R^2\theta^2_j/d^2_0\right),
\end{equation}
where the position of the reference particle has been set to $\theta_i=0$ without loss of generality, due to rotational symmetry, thereby making it independent of the reference index $i$. This also ensures that the interparticle distance lies between $-L/2$ and $L/2$, thereby satisfying the minimum image convention. Substituting $y=R\theta_j/d_0$, the integral takes the form
\begin{equation}
    \mathscr{A}_i=2 d_0 \int_{0}^{L/2d_0} \D y\ \exp\left(-y^2\right)=d_0\sqrt{\pi}\ E\left(L/2d_0\right),
\end{equation}
where $E(x)$ is the error function.
Therefore, the required integral is obtained as
\begin{equation}
I_i=\mathscr{A}_i^2=\pi d^2_0 E^2\left(L/2d_0\right),
\end{equation}
which can then be used to obtain the effective coupling constants in Eqs.~\eqref{compact-form}-\eqref{effective-strength}.

\section{Effect of ensemble and domain size}
\label{sys-size}

\begin{figure}[b]
    \centering
    \subfloat[$N=64$, $128$, and $256$ (left to right).\label{N_vary_a}]{\includegraphics[width=0.33\linewidth]{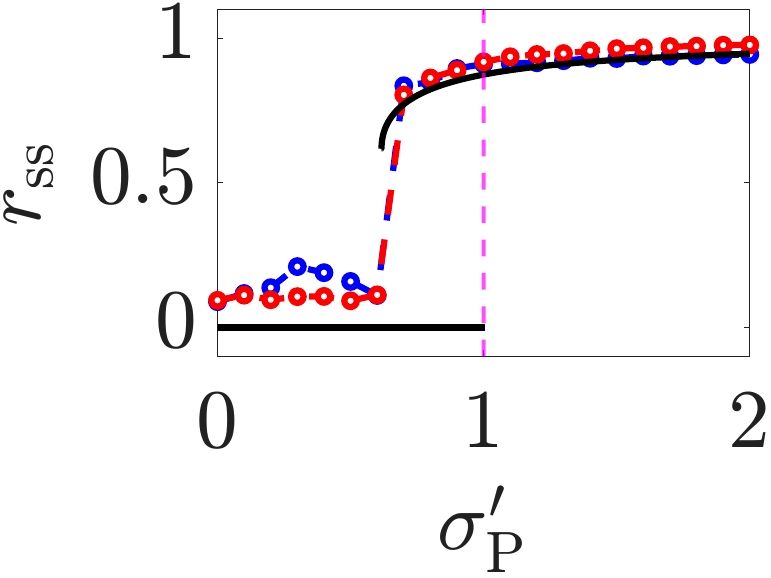}\includegraphics[width=0.33\linewidth]{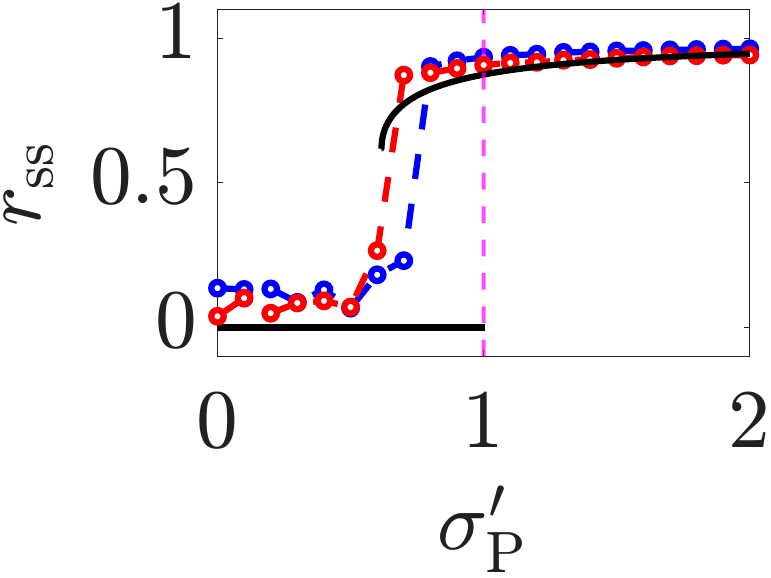}\includegraphics[width=0.33\linewidth]{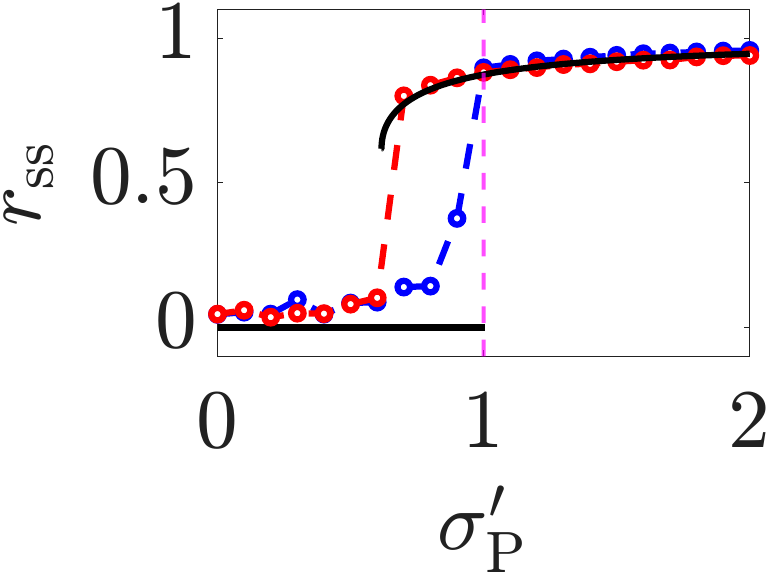}}

    \subfloat[$N=512$, $1024$, and $4096$ (left to right).\label{N_vary_b}]{\includegraphics[width=0.33\linewidth]{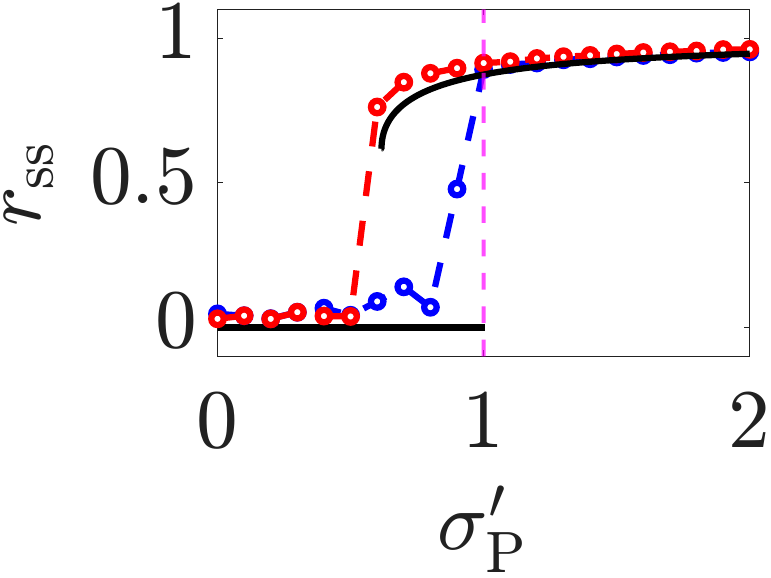}\includegraphics[width=0.33\linewidth]{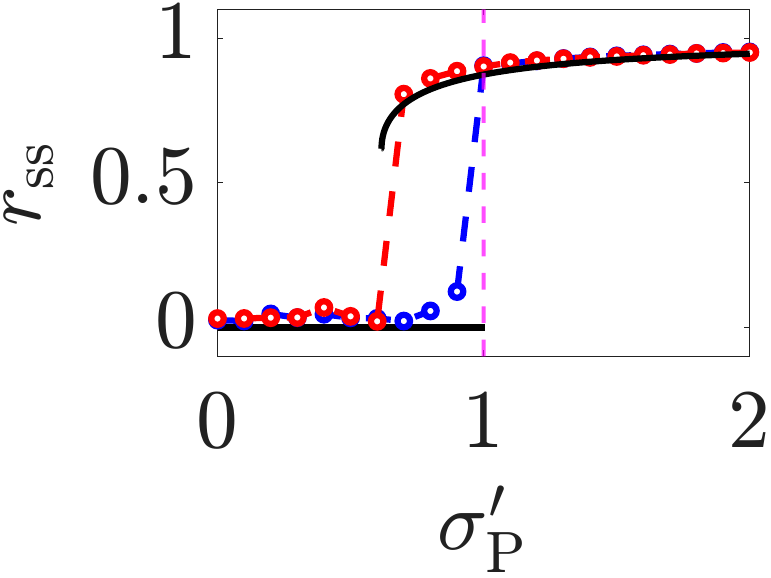}\includegraphics[width=0.33\linewidth]{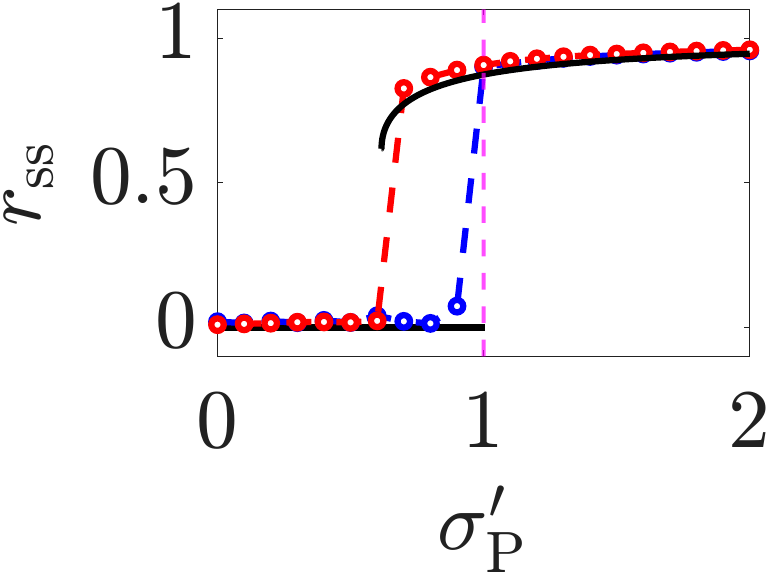}}

    \subfloat[$L=4$, $8$, and $16$ (left to right).\label{L_vary_a}]{\includegraphics[width=0.33\linewidth]{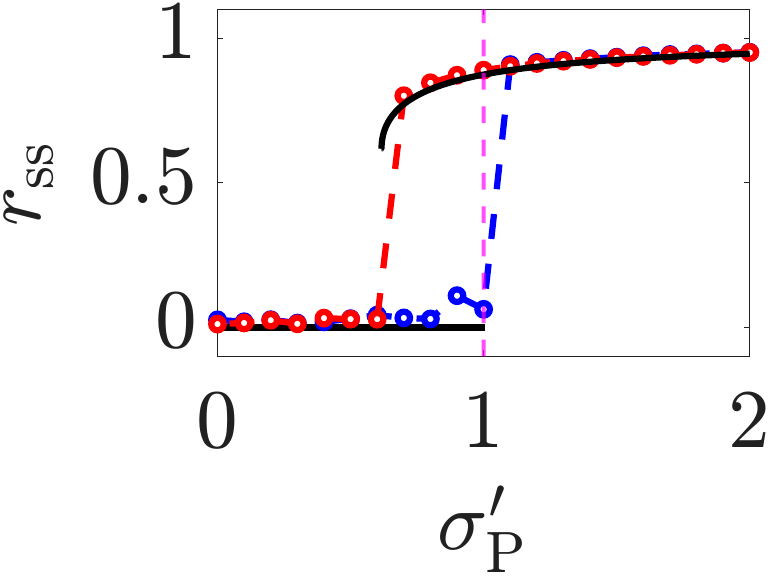}\includegraphics[width=0.33\linewidth]{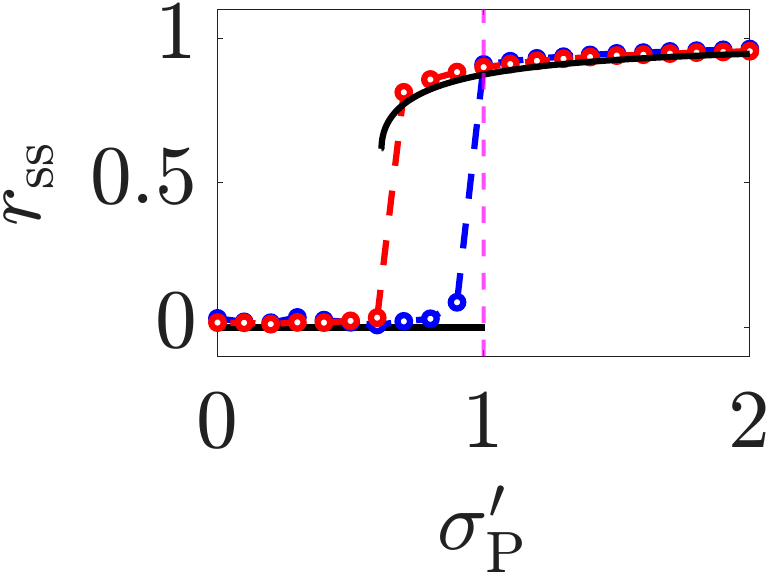}\includegraphics[width=0.33\linewidth]{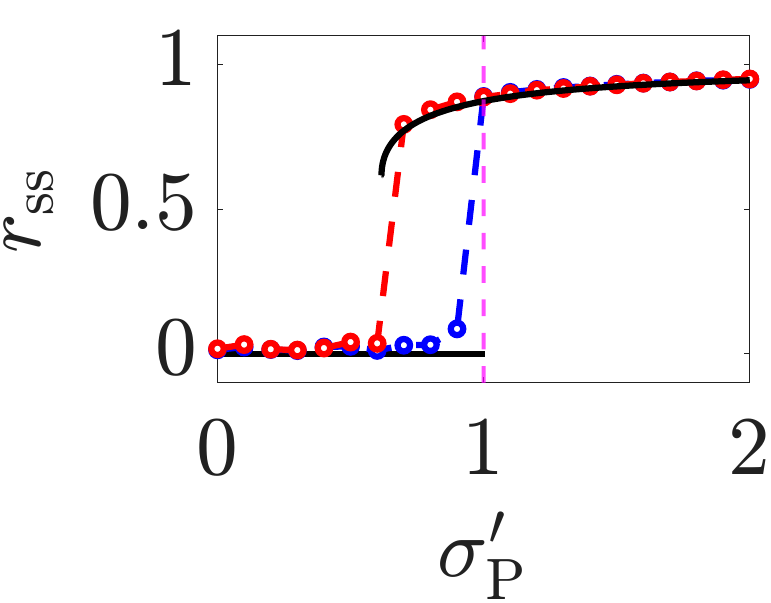}}

    \subfloat[$L=64$, $128$, and $256$ (left to right).\label{L_vary_b}]{\includegraphics[width=0.33\linewidth]{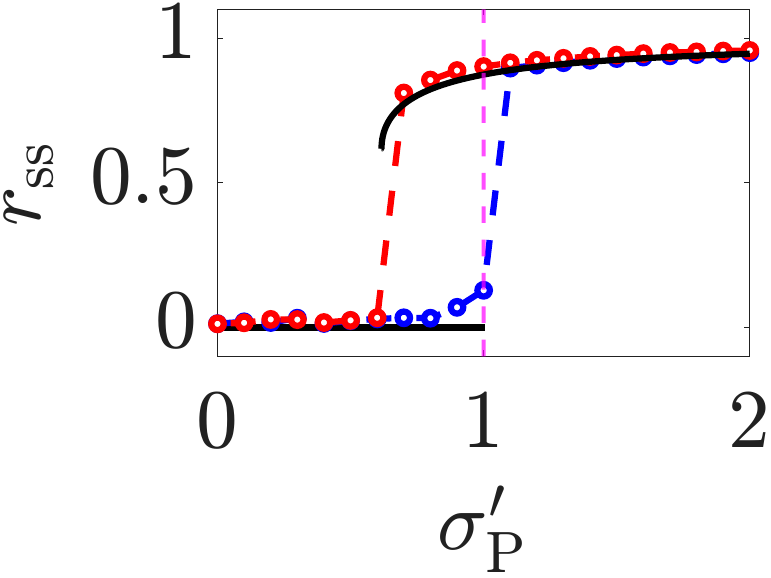}\includegraphics[width=0.33\linewidth]{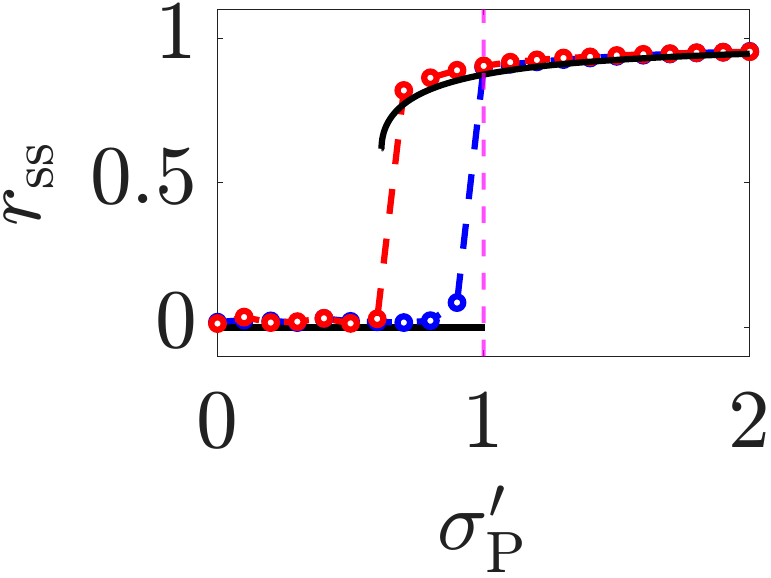}\includegraphics[width=0.33\linewidth]{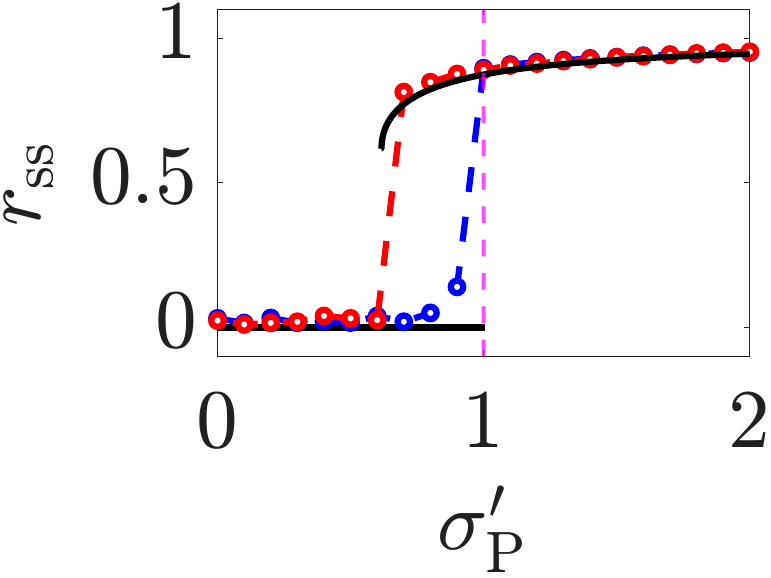}}

    \caption{$r_{\rm ss}$ as a function of $\sigma^{\prime}_{\rm P}$ for different values of $N$ (in (a) and (b)) and $L$ (in (c) and (d)). In the former, $L=32$, whereas $N=2048$ in the latter. Here, the values of $(T_{\rm R},T_{\rm C})$, the meaning of the blue and red curves, as well as the vertical line, are identical to Fig.~\ref{phase-transition-common}, whereas $(k,\Omega,\gamma,D,\sigma_{\rm R})$ are identical to Fig.~\ref{v0_1_c_vary}.}
    \label{size_vary}
\end{figure}

It is worth mentioning that the analytical results in Sec.~\ref{math} are derived for an infinitely large ensemble of agents. However, computer simulations are run for a finite-sized system and need to be optimized to capture the essential features reliably. To this end, the curves in Fig.~\ref{size_vary} demonstrate the phase transitions for different values of either $N$ or $L$, while keeping the other parameter constant. In particular, Figs.~\ref{N_vary_a} and \ref{N_vary_b} clearly indicate that, for lower ensemble sizes (\ie, $N=64$, $128$, and $256$), the asynchronous branch of the bistable region is not captured accurately; this can be attributed to finite-size fluctuations, which drive the system away from the asynchronous state. However, there is also no significant improvement in accuracy changing the system size to $N=512$, $1024$, or $4096$.
In contrast, upon varying $L$ while keeping $N$ constant, there is no qualitative change in the phase transitions, and the agreement between theoretical and numerical simulations persists; see Figs.~\ref{L_vary_a} and \ref{L_vary_b}. Consequently, the choice $N=2048$ balances both accuracy and computational time, whereas $L=32$ is chosen in an ad hoc manner such that $\rho=2.0$; this combination is used throughout the article.

\bibliography{aipsamp}

\end{document}